\documentclass[10pt,journal,compsoc]{IEEEtran}

\usepackage{microtype}                 
\PassOptionsToPackage{warn}{textcomp}  
\usepackage{textcomp}                  
\usepackage{mathptmx}                  
\usepackage{times}                     
\usepackage{cite}                      
\usepackage{booktabs}                  
\usepackage{amsmath}
\usepackage{amssymb}
\usepackage{amsfonts}
\usepackage{multirow}
\usepackage[table,xcdraw]{xcolor}
\usepackage{enumitem}
\usepackage{graphicx}
\usepackage[colorlinks,bookmarksopen,bookmarksnumbered,allcolors=black]{hyperref}
\usepackage{subfig}

\definecolor{Goldenrod}{RGB}{255,223,67}
\definecolor{SpringGreen}{RGB}{198,220,102}
\definecolor{Peach}{RGB}{247,149,90}

\newif\ifnotes
\notestrue

\newcommand{\changed}[1]{#1}

\begin{document}


\title{Fitting Bell Curves to Data Distributions\\ using Visualization}

\author{Eric Newburger, Michael Correll, and Niklas Elmqvist,~\IEEEmembership{Senior Member,~IEEE}
\IEEEcompsocitemizethanks{
    \IEEEcompsocthanksitem Eric Newburger and Niklas Elmqvist are with University of Maryland, College Park, MD, United States. E-mail: enewburg@terpmail.umd.edu, elm@umd.edu 
    \IEEEcompsocthanksitem Michael Correll is with Tableau Research, Seattle, WA, United States. E-mail: mcorrell@tableau.com}
    \thanks{Manuscript received XXX XX, 2021; revised XXX XX, 2021.}
}

\markboth{IEEE Transactions on Visualization and Computer Graphics}{Newburger \MakeLowercase{\textit{et al.}}: For Whom the Bell Tolls}

\IEEEtitleabstractindextext{%
\begin{abstract}
    Idealized probability distributions, such as normal or other curves, lie at the root of confirmatory statistical tests.
    But how well do people understand these idealized curves?
    In practical terms, does the human visual system allow us to match sample data distributions with hypothesized population distributions from which those samples might have been drawn?
    And how do different visualization techniques impact this capability?
    This paper shares the results of a crowdsourced experiment that tested the ability of respondents to fit normal curves to four different data distribution visualizations: bar histograms, dotplot histograms, strip plots, and boxplots.
    We find that the crowd can estimate the center (mean) of a distribution with some success and little bias.
    We also find that people generally overestimate the standard deviation---which we dub the ``umbrella effect'' because people tend to want to cover the whole distribution using the curve, as if sheltering it from the heavens above---and that strip plots yield the best accuracy.
\end{abstract}

}

\maketitle
\IEEEdisplaynontitleabstractindextext
\IEEEpeerreviewmaketitle

\begin{figure*}[htb]
    \centering
    \subfloat[Bar histogram.]{
        \resizebox{0.4\textwidth}{!}{\includegraphics{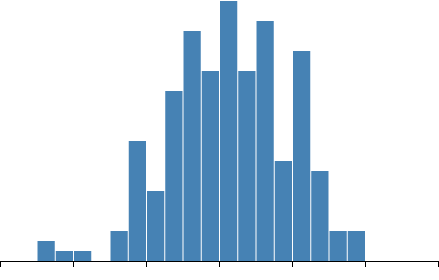}}
        \label{fig:stimulus-bar}
    }
    \subfloat[Wilkinson dot plot.]{
        \resizebox{0.4\textwidth}{!}{\includegraphics{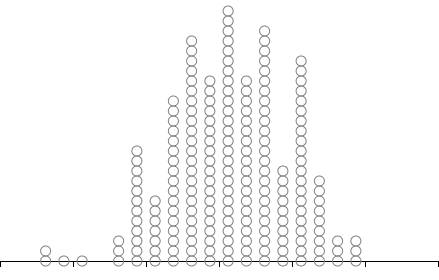}}
        \label{fig:stimulus-dot}
    }\\
    \subfloat[Boxplot.]{
        \resizebox{0.4\textwidth}{!}{\includegraphics{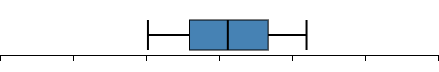}}
        \label{fig:stimulus-box}
    }
    \subfloat[Strip plot.]{
        \resizebox{0.4\textwidth}{!}{\includegraphics{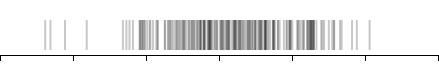}}
        \label{fig:stimulus-strip}
    }
    \caption{\textbf{Experimental setup.} 
    One-dimensional dataset visualized using the four different visual representations that we studied in our experiment. 
    In the experiment, participants fitted a normal curve on these representations.}
    \label{fig:teaser}
\end{figure*}

\section{Introduction}

\IEEEPARstart{M}{any} crucial visualization tasks rely on not just an assessment of individual values, but a holistic assessment of the overall \textit{distribution}: identifying points as outliers, judging variability, assessing the appropriateness of various parameterized statistical tests, or even just building up a picture of likely and unlikely values: all require fitting, implicitly or explicitly, distributions to sample data.
While prior work has examined aggregate or ensemble graphical perception tasks~\cite{szafir2016}, the specific fitting of curves of idealized distributions we believe is both understudied and has the potential to inform the design of statistical graphics, especially for novice users.
We choose the fitting of normal distributions to sample data as a graphical perception ``fruit fly''~\cite{Rensink2014}: a relatively simple and well-understood problem that is nonetheless important for a variety of tasks in statistics.

Our work stems from an open question in visualization: the extent to which human visual judgments from statistical graphics can operate as a sort of ``visual statistics''~\cite{Correll2015} or ``graphical inference''~\cite{Buja2009}: that is, the extent that we can rely on our ability to read or aggregate information in charts to assess effect sizes, provide evidence against null hypotheses, or assess trends in our data.
If these are abilities are robust, they point to the possibility to augment or even replace formal statistical tests with visual appraisals.
Yet, there is also evidence that in some ways the visual estimation of statistical properties is insufficient, non-anologous, or otherwise disconnected from the results of formal statistical tests.
Bias~\cite{Correll2014}, satisficing strategies~\cite{Kale2021}, and perceptual proxies~\cite{DBLP:journals/tvcg/OndovYKEF21,Yuan2019} can produce mismatches between statistical assessments of data and human judgments or estimates.
It therefore is necessary to perform empirical assessment of human performance and measurement of human biases in reading statistical graphics before promoting their use as inferential or confirmatory tools.

In this paper, we present results from a preregistered and crowdsourced user study investigating how well members of the general population are able to fit normal curves to data distributions when represented in a variety of graphic forms.
For each trial, participants were shown a visualization of the data and were asked to move the centerpoint (mean) and width (spread or standard deviation) of a Gaussian curve overlaying the sample to create the best possible fit.
The visualizations studied included bar histograms, Wilkinson dotplot histograms, strip plots, and boxplots (Figure~\ref{fig:teaser}).

\section{Related Work}
\label{sec:related-work}

Beyond its use for communication, visualization is often touted as a tool primarily for \textit{exploratory data analysis}~\cite{Tukey1977} due to its investigative and data-driven nature. 
However, visualization can also be used for some forms of \textit{confirmatory data analysis}~\cite{Lehmann2005, Schervish1995}, at least as a complement.
We review these topics in detail.

\subsection{Graphical Inference}

Creating graphical representations of data is a common and natural part of statistical workflows~\cite{Cleveland1993}, and even central to some, such as exploratory data analysis~\cite{Tukey1977}.
Accordingly, making inferences from these graphical representations---i.e., \textit{graphical inference}---is a commonplace complement to formal statistical tests.
Early examples of such practice date back to Scott et al.~\cite{Scott1954} validating astronomical models by generating artificial star charts using model parameters and then asking people to compare them to real charts.

However, it is only recently that the community has begun to ask how graphical representations of data can support higher-order tasks beyond merely reading values, trends, and outliers.
Buja et al.~\cite{Buja2009} proposed frameworks for \textit{visual statistics}, where the visual representations provide the test statistic and human cognition is the statistical test.
They demonstrate this approach using a ``Rohrschach'' test of random data, as well as a lineup of small multiples, only one of which uses real data.
In followup work, Wickham et al.~\cite{Wickham2010} adapted the idea to the visualization community, describing how these protocols can be used with common visualizations to uncover new findings while avoiding false positives.
Beecham et al.~\cite{Beecham2017} applied this ``lineup'' protocol for graphical inference to geographic clustering visualizations.
Correll et al.~\cite{Correll2019} used it to investigate whether common distribution graphics were effective in displaying outliers or gaps.

Research has also been conducted on how well people can retrieve aggregate statistics from visualizations.
Correll et al.~\cite{Correll2012} studied how line graphs and other visualizations can be designed to enable accurate comparisons of averages in time-series data.
Albers et al.~\cite{Albers2014} generalized this idea to six aggregate tasks for eight different time-series visualizations.
Furthermore, Aigner et al.~\cite{journals/cgf/AignerRH12} enriched line graphs with color to better support visual statistics, and Fuchs et al.~\cite{conf/chi/FuchsFMBI13} derived line glyphs to support higher-level aggregate tasks.
Correll and Heer~\cite{conf/chi/CorrellH17} studied people's ability to fit trend lines to bivariate visualizations in a crowdsourced experiment.

Deriving aggregate statistics from other visual representations is also relevant to our work.
Fouriezos et al.~\cite{Fouriezos2008} asked participants to compare the average height of two groups of bar charts, yielding high accuracy improved by the number of bars, but impaired by variance.
Gleicher et al.~\cite{Gleicher2013} study mean value judgments in multi-class scatterplots, finding that performance is reliably high independent of the number of points and conflicting encodings.
Based on results from crowdsourced experiments, Correll and Gleicher~\cite{Correll2014} propose redesigns of error bars in bar charts, showing how violin or gradient plots produce insights more aligned with statistical inference. 
Finally, Nguyen et al.~\cite{Nguyen2020} used a crowdsourced experiment to explore how different visual aggregations might impact users' perception of summary statements about sample populations.
They found no impact of visual aggregation strategy on participant accuracy, but noted that participants that were shown the full data were less confident and less prone to engaging in dichotomous thinking.
This suggests that dis-aggregated visualizations give a richer and more nuanced view of the underlying data.

\subsection{Ensemble Processing}

A central aspect of our research is how people visually estimate set characteristics in a visualization. 
A common trend in vision research is to use abstract dot clusters because these allow for precisely controlling the visual stimulus~\cite{Whitaker1988}.
Morgan and Glennerster~\cite{Morgan1991} studied perception of centroids in such clusters and found very high accuracy---with some individual differences.
Ariely confirmed these results, finding that people generally are quite accurate in estimating means and ranges in a set of points even if they quickly forget specific details~\cite{Ariely2001}.
This suggests that the visual system extracts statistical rather than individual details of sets.

Naturally, human perception may behave differently for geometric shapes than for random dot clusters.
Melcher and Kowler~\cite{Melcher1999} study eye movements (saccades) during centroid estimation on shapes formed using outlines created by such dots. 
They found that people in general are quite proficient at establishing the centerpoint of a target shape's area, even when presented with skewed distributions of dots as well as distracting clusters.
This suggests that the shape itself---as defined by its outline---guides perception rather than individual point primitives.

But how do we from many individual dots to entire shapes when our perceptual system is limited to perceiving only a handful of entities?
The answer may lie in \textit{ensemble processing}~\cite{Alvarez2011}, which deals with how the visual system computes averages of many types of visual features to handle complex shapes and configurations.
This is easier for actual objects in the real world, which tend to have regular shapes, than for artificial dot clusters.
This kind of computing has also been known as \textit{perceptual averaging.}
Chong and Treisman~\cite{Chong2003} presented results from an experiment showing evidence that participants conducted size averaging even for clusters of 12 shapes.
Albrecht and Scholl~\cite{Albrecht2010} extended these results to dynamically changing displays with similar outcomes.

Some work exists on studying these phenomena in visualization.
Most relevant to our work, Szafir et al.~\cite{szafir2016} show how ensemble processing can support a wider variety of relevant tasks in statistical graphics, including \textit{summarization} of values and \textit{estimation} of structures. As the visual complexity rises, however, precision often decreases.
Yuan et al.~\cite{Yuan2019} found that estimating graphical attributes of shapes in a visualization during multi-value comparison often reduces to primitive perceptual cues, yielding lower accuracy than when perceiving single values.
These perceptual cues, or \textit{proxies}~\cite{Yuan2019, DBLP:journals/tvcg/OndovYKEF21}, provide shortcuts for when the visual system is asked to compare multiple shapes in parallel. 
In contrast, our work in this paper focuses on a more holistic task of fitting a mean and spread of an idealized bell curve to a single visualization of data distributions.

\subsection{Visualizing Distributions}

One of the most common visualizations for univariate distributions is the histogram, which aggregates data occurrences into discrete ranges (``bins'') and visualizes the resulting counts using bars.
However, histograms have flaws, most of them related to bin size and bin number~\cite{Correll2019}.
Alternate representations include strip plots, density plots, violin plots, and gradient plots~\cite{conf/chi/KayKHM16}, but these lack the easy familiarity of histograms.
Even disregarding binning aspects, the aggregating nature of histograms can both be a strength and a weakness: a strength, because the representation is robust in visualizing large datasets, but a weakness because the bars convey information about the relative, rather than the absolute, number of cases in each bin.
Interested users must look to the axis for absolute counts---a reading task rather than a mere seeing task.
For this reason, Wilkinson dot-histograms~\cite{Moon2016}, where each discrete item in a bin is represented as a circle in a stack of circles, may improve on the traditional bar-based design~\cite{Correll2019, Hullman2018}.

Uncertainty visualization is a closely related topic, since we often calculate uncertainty as an idealized distribution of potential variance around a point estimate, and significant progress has been made recently on designing visualization techniques where the uncertainty is intrinsic to the representation.
In one example, this approach yielded a significant confidence improvement when estimating outcomes using a so-called quantile dotplot~\cite{conf/chi/KayKHM16}.
Hypothetical outcome plots (HOPs)~\cite{Hullman2015} use animated draws to illustrate uncertainty, and have shown superior performance compared to violin plots and error bars.

Finally, our methodology in this paper is inspired by recent work in uncertainty visualization that uses \textit{graphic elicitation}~\cite{Crilly2006} by asking participants to draw visual representations of data as part of the evaluation.
For one thing, doing this improves recall and comprehension; Kim et al.~\cite{DBLP:conf/chi/KimRH17} found that graphically eliciting visual forms of a participant's prior knowledge and observed data helped them remember and reason about this.
In follow-up work, Hullman et al.~\cite{Hullman2018} asked participants to sketch their predictions of uncertainty distributions using both continuous and discrete representations prior to seeing the actual distributions.
Their findings show that participant predictions are also improved by such graphic elicitation.
Finally, Kim et al.~\cite{DBLP:conf/chi/KimWKH19} elicit prior and posterior beliefs for participants to derive a Bayesian cognition model for how people interpret data visualizations.
Our work uses similar elicitation methods to ask participants to fit a graphical representation of a continuous curve to visualizations, but since our distribution is Gaussian, we ask only for mean and spread.

\section{Study: Fitting Bell Curves}

The goal of our study was to understand how well, or even if, untrained people would be able to fit normal curves to a data sample drawn from a normal distribution.
To begin exploring this question, we conducted a crowdsourced user study on Amazon Mechanical Turk where participants were asked to control the position (mean, or ``center point'') and width (standard deviation, or ``spread'') of a Gaussian curve to fit a data sample.
Samples were represented on screen using four different visualizations: a bar histogram, a dot histogram, a strip plot, and a boxplot.
The variety of representations made it less likely that the characteristics of any one distribution graph type would bias the experimental results, and also allowed for the possibility of deriving design recommendations for creating effective data distribution graphics. 
Our design varied the size of the random sample ($n = 50$ or $200$), the noise in the data (coefficients of variation = 0.2 or 0.4), and the visual representation of the data samples.
To minimize the possibility of a ``left-to-right bias,''~\cite{Spalek2005} where respondents get into a habit of always moving the curve from left to right on the screen, the means of some data samples were adjusted to move their center points to the left on the screen (values below zero).
These moves did not affect the shape of the data, and coefficient of variation (CV) calculations were based upon the original positions of datasets. 

An original version of this study only included Wilkinson dot histograms, and was preregistrered on OSF.\footnote{\url{https://osf.io/behwz}}
Since that initial study, we have expanded our experiment to include three additional visual representations: traditional bar histograms, strip plots, and boxplots.
The new preregistration can be found on OSF.\footnote{\url{https://osf.io/a9b48}}
The discussion below only concerns the new experiment; the original data for 200 participants are not included in this paper and are thus not reported.
Below we review our methods, followed by our results next.

\begin{figure*}[tbh]
    \centering
    \resizebox{\textwidth}{!}{\includegraphics{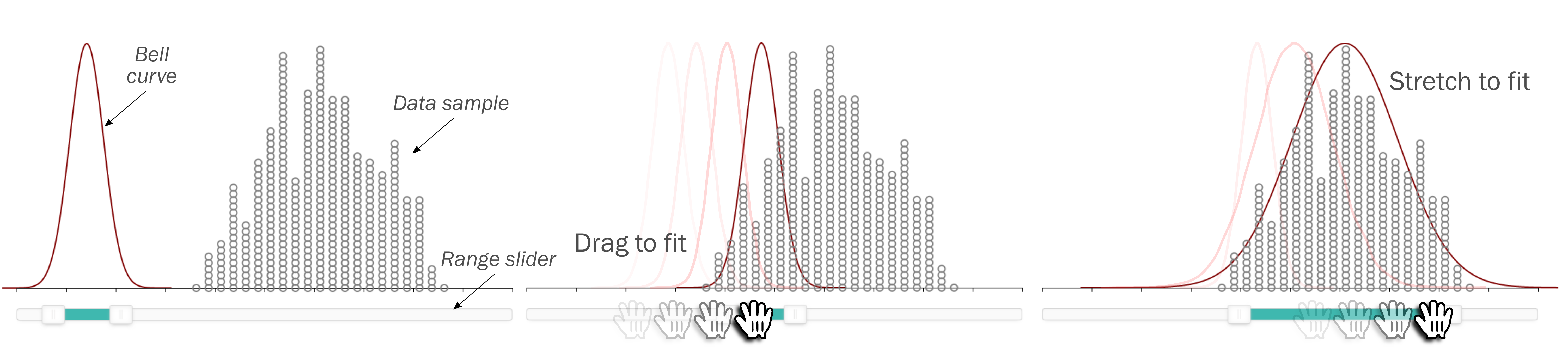}}
    \caption{\textbf{Curve fitting task.}
    Typical sequence in our crowdsourced curve fitting experiment.
    Participants controlled a normal curve using a range slider.
    They fit this continuous curve on top of a data sample (here represented by a Wilkinson dotplot histogram).
    Our evaluation varied the visualizations as well as the number of data points and the coefficient of variation for samples.
    }
    \label{fig:setup}
\end{figure*}

\subsection{Participants}

Because this study focused on low-level perceptual tasks that require no specific training or prior data visualization expertise, we conducted our study using Amazon Mechanical Turk (MTurk).
While the use of MTurk means that we have little control over participant demographics and expertise as well as their computer hardware, prior work has shown that simple visual tasks are particularly amenable to this kind of crowdsourced study~\cite{Heer2010}.

In our experiments, all experimental factors were within-participants. 
We planned to recruit a total of 100 participants.
We limited participation to the United States due to tax and compensation restrictions imposed by our IRB.
We screened participants to ensure at least a working knowledge of English; this was required to follow the instructions and task descriptions in our testing platform.
Participants were prevented from participating in the experiment more than once.
All participants were ethically compensated at a rate consistent with an hourly wage of at least \$10/hour (the U.S.\ federal minimum wage in 2020 was \$7.25).
More specifically, the payout was \$2.50 per session, and with a typical completion time of 14 minutes and 54 seconds, this yielded an hourly wage of approximately \$10.00/hour.

\subsection{Apparatus}

Because of our crowdsourced setting, we were unable to control the specific computer equipment that the participants used.
The study was distributed through the user's web browser.
We required that all devices were personal computers (laptop or desktop) or touch tablets; smartphones were disallowed due to the limited screen space available on such devices.
We also required participants to use browser windows of at least $1280 \times 800$ pixels.

\begin{figure}[tbh]
    \centering
    \includegraphics[width=\columnwidth]{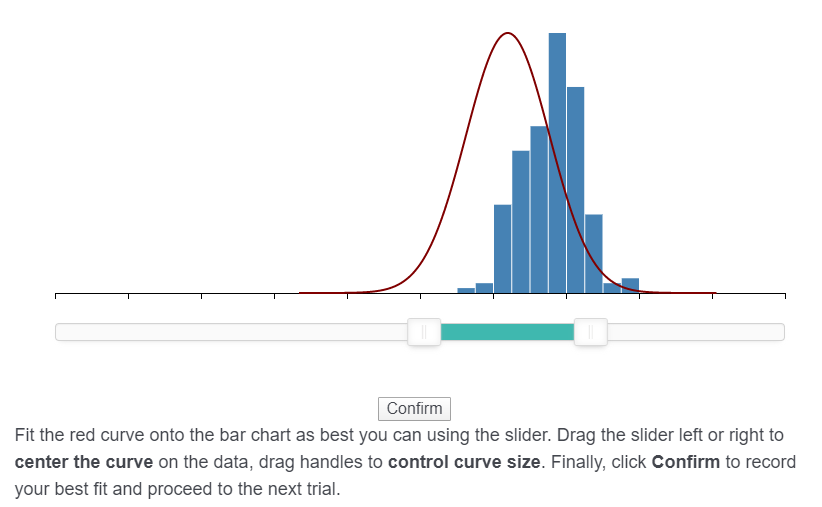}
    \caption{\textbf{Curve fitting task (bar histogram).}
    The red Gaussian curve is controlled by the user.
    The bar histogram represents the sample to fit. 
    The range slider below the axis controls the spread using the width of the range, and the centerpoint using its position on the slider.
    Once the participant is satisfied with the fit, they click the ``Confirm'' button to finish and then proceed to the next trial.
    }
    \label{fig:task}
\end{figure}

\subsection{Task}

\begin{figure}[htb]
    \centering
    \includegraphics[width=\columnwidth]{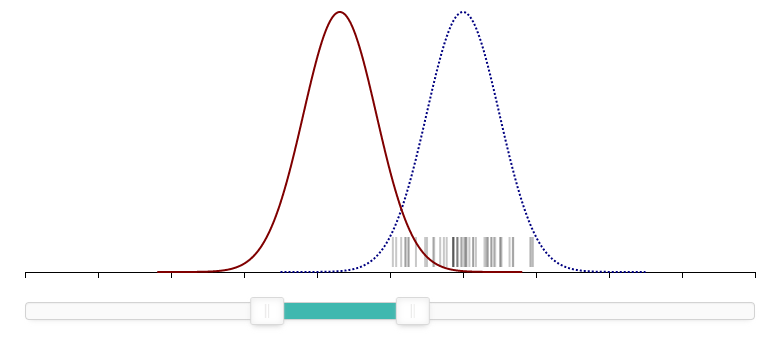}
    \caption{\textbf{Training trial.}
    Participants were asked to match the curve (red line) with a correctly fitted curve (blue line).
    This example shows a strip plot; we included training trials for all visualizations.
    These trials also served as attention trials for our experiment.
    }
    \label{fig:training}
\end{figure}

The tasks consisted of fitting a normal (Gaussian) curve onto a data visualization using a range slider~\cite{Ahlberg1992} that controlled the spread (i.e., standard deviation) of the curve using the width of the interval and its center point (i.e., mean) by moving the position of the interval on the slider.
Figure~\ref{fig:task} shows a screenshot of a typical task.
Participants were instructed to find the ``best fit'' between the curve and the visualized data.
In a training trial for each visualization block (which we also used as attention trials; see below for details), participants were shown a perfect fit using a curve with a contrasting color, and were asked to match their own curve with the correct answer (Figure~\ref{fig:training}).

The testing platform was implemented in JavaScript using D3~\cite{Bostock2011} and embedded into a Qualtrics survey accessed using the participant's web browser.
We used \texttt{noUiSlider}\footnote{\url{https://refreshless.com/nouislider/}} for the range slider implementation.
Participants moved the center of the range slider both by moving one of the range endpoints, or by dragging the center area of the range to move both endpoints simultaneously.
In other words, participants were able to directly control both the spread and the mean of the curve.
However, participants were unable to drag the visual representation of the curve itself.

\subsection{Dataset Generation}

Data for all trials was controlled so that all participants saw the same datasets during a session.
All datasets were randomly drawn from a normal distribution with varying degrees of spread, and then iteratively jittered until the standard deviation fell within 1\% of the intended values for the spread $S$ (see below).
Datasets were generated so that they fell within $[-3, 7]$, with the horizontal axis fixed at $[-10, 10]$.
However, no value labels were shown for the axis to minimize number bias.
The histogram used 50 bins across the horizontal axis, but actual trials typically used between 4 and 8 bins in total (based on spread).

We chose not to introduce any datasets drawn from any other distribution than the normal one, as our purpose with this experiment is to fit normal curves rather than having participants determine which distribution to use.
Regardless, by varying the experimental factors (below), we are able to get distributions that are sufficiently noisy in appearance.
However, it means that our datasets were all more or less symmetric.

\subsection{Experimental Factors}

We chose to model three factors in our experiment:

\begin{itemize}
    \item\textbf{Data Size} ($D$): The number of samples in the dataset being fitted.
    As the number of samples increases, a histogram will begin to approach the idealized distribution. 
    We chose two levels: 50 and 200 samples.
    While people may want to fit smaller sample sizes in practice, we chose 50 items as the smallest level because the Student t-distribution~\cite{Student1908}---which we are interested in supporting using graphical inference---only begins to conform to the normal distribution for larger samples (30 items or more).
    
    \item\textbf{Spread} ($S$): The standard deviation of the Gaussian distribution being fitted.
    We chose two levels for this factor expressed as the \textit{coefficient of variation} (CV) (or relative standard deviation), i.e., as a ratio between the standard deviation $\sigma$ and the mean $\mu$ ($\sigma/\mu$): 20\% and 40\%.
    We settled on these values through pilot testing to ensure that this yielded both relatively noisy (for high values of $S$) as well as relatively tight (for low values of $S$) samples.
    Because our data generation is stochastic, datasets are only guaranteed to have specific spreads within a tolerance of 1\%. 
    
    \item\textbf{Visualization} ($V$): The visualization type used to represent the dataset.
    Based on our review of the literature, we chose four distinctive visualization techniques (Figure~\ref{fig:teaser}):
    \begin{itemize}
        \item\textbf{Bar histogram:} A ``classic'' histogram where the aggregated number of data items for each bin is represented using a bar of uniform width (Figure~\ref{fig:stimulus-bar}).
        
        \item\textbf{Wilkinson dot histogram:} A variant of histograms initially proposed by Wilkinson~\cite{Wilkinson1999dotplot}, dot histograms are unit visualizations~\cite{Park2018b} that organize  individual dots (circles) for each item into bars for each bin (Figure~\ref{fig:stimulus-dot}).
        
        \item\textbf{Boxplot:} A box-and-whisker plot as pioneered by John W.\ Tukey~\cite{wickham201140}, where a central rectangle contains the middle half of the data (from the 25$^{th}$ to the 75$^{th}$ percentile), the median (50$^{th}$ percentile) is marked with a line, and the ``whiskers'' mark borders of wider percentiles, in this case the upper 10\% and lower 10\% of the data (the 10$^{th}$ and 90$^{th}$ percentiles) (Figure~\ref{fig:stimulus-box}).
        Boxplots in this study were \textbf{not} augmented with icons like dots or stars to indicate outliers.
        
        \item\textbf{Strip plot:} A unit visualization~\cite{Park2018b} where each item is drawn as a short vertical line with opacity on the horizontal axes (i.e., with no vertical data encoding), yielding a representation similar to a barcode (Figure~\ref{fig:stimulus-strip}).
    \end{itemize}
    
\end{itemize}

While other visual representations exist, we felt this to be a representative selection on the spectrum of data aggregation: strip plots draw all values, bar and dot histograms bin them, and boxplots discard them all in favor of descriptive statistics on the sample.

Furthermore, the number of bins is an important parameter for histograms~\cite{Correll2019}.
We opted not to model this factor directly, instead keeping the extents of the horizontal axis constant (at $[-10, 10]$) and the number of bins constant (50), thus indirectly controlling the number of bins used depending on spread $S$.

\subsection{Experimental Design}

We used a within-participant design, where each participant saw all data sizes, spreads, and visualizations. 
The relatively small total number of conditions enabled us to keep sessions shorter than 20 minutes in duration to minimize fatigue and maximize attention for crowdworkers.
The order of trials was randomized for each individual participant.
This yielded the following design:

\bigskip

\begin{tabular}{crl}
  & 2 & \textbf{Data Size} $D$ (50, 200 samples)\\
  $\times$ & 2 & \textbf{Spread} $S$ (0.2, 0.4)\\
  $\times$ & 4 & \textbf{Visualization} $V$ (bar, dot, box, strip)\\
  $\times$ & 3 & repetitions\\
  
  \hline

  & 48 & trials per participant\\

\end{tabular}

\bigskip

For 100 participants, we planned to collect a total of 4,800 trials.\footnote{Note that due to a miscommunication within the research team, our preregistration called for 100 participants but we ended up recruiting 150.
We discuss this later in our section on deviations from the preregistration.}
For each trial, we captured the completion time as well as the accuracy.
The completion time was measured from when the trial was displayed to the participant until the participant submitted an answer.
The accuracy measure was based on two metrics:

\begin{itemize}

\item\textbf{Mean error:} If $\mu_c$ is the mean of the normal curve fitted by the participant and $\mu_s$ is the actual mean of the sample, we calculate the mean error as $\mu_c - \mu_s$.
In most of our analysis, we take the absolute value of this metric.

\item\textbf{Standard deviation error (\%):} If $\sigma_c$ is the standard deviation of the normal curve fitted by the participant and $\sigma_s$ is the actual standard deviation of the sample, we calculate the standard deviation error as $(\sigma_c - \sigma_s) / \sigma_s$, expressed as a percentage (since this measure is independent of screen resolution, a ratio provides more information).
In most of our analysis, we take the absolute value of this metric.
\end{itemize}

\subsection{Procedure}

All recruitment was conducted via Amazon Mechanical Turk.
Participants that fit the eligibility criteria opened the survey in a separate browser window.
At the end of their participation, they copied a unique completion code back into the Mechanical Turk interface, and were later paid as their work was checked. 

Each session started with a consent form with waived signed consent. 
Failing to give consent terminated the experiment.
Participants were instructed that they could abandon their session at any point in time.
Unfortunately, we were unable to pay participants who only completed a partial session.
We informed participants of this fact in the consent form when starting the session.

After consenting, participants were asked demographic questions about their age, education level, and knowledge of statistical concepts.
We also asked participants to reaffirm that they were using a tablet or computer to participate.
Then participants were shown practice trials for each visualization type where they were given instructions on how to read the visualization and complete the task. 
For each such practice trial, a correctly fitted curve was shown in a contrasting blue color. 
These practice trials also served as ``attention trials.''
The purpose of these attention trials was to eliminate responses from crowdworkers who did not pay attention to the task.
Any session where the participant responded with an error of more than 3 standard deviations from the actual mean for these attention trials were discarded from analysis.
We included information about this fact in the consent form.

Each individual trial started with the display of the dataset and the curve (Figure~\ref{fig:task}) and ended when participants clicked the ``confirm'' button.
Participants were unable to confirm a trial before interacting at least once with the range slider.
Completion time was measured from the display of the trial, to this button-click.
Participants were instructed to use the intermission between visualization blocks if they needed to rest between trials.
A progress bar at the top of the screen showed the study progress.

Typical sessions lasted between 14 and 15 minutes.
A few participants used much more time to complete their sessions, but our logs indicate that these participants took long breaks between trials (presumably due to interruptions).
We believe that the effective time spent on the experiment was less than 15 minutes.

\subsection{Hypotheses}
\label{sec:hypotheses}

We preregistered the following hypotheses about our experiment:
 
\begin{itemize}

\item\textit{Estimation of means will be more accurate than estimation of spread.} 
Our intuition from our own experience as well as our literature review (e.g., that people are able to visually determine averages with high accuracy)~\cite{Gleicher2013} is that people will be capable of determining the average with high accuracy, but fitting the curve over the sample will be less accurate.

\item\textit{Participants will be more accurate at estimating both mean and spread as the number of points increases.}
For larger datasets, the impact of sampling error will be lower and the overall shape of the distribution more well-defined.
This should make it easier to perceptually estimate the mean.

\item\textit{There will be non-uniformity in performance across visualization types.}
In particular, we anticipate that:
\begin{itemize}
    \item\textit{Participants will be more accurate at estimating means with boxplots.}
    Boxplots directly encode the median of the distribution in its visual representation, which is close to or identical to the mean in most of our stimuli.
    
    \item\textit{Participants will be less accurate at estimating means with strip plots.}
    The use of opacity and the impact of overplotting to encode density makes precise estimation difficult.
\end{itemize}

\item\textit{Participants will consistently overestimate the spread of distributions, except for in boxplots.}
Based on prior work suggesting the use of ``perceptual proxies'' for ensemble processing and visual statistical tasks~\cite{DBLP:journals/tvcg/OndovYKEF21}, we anticipate that participants will attempt to cover all of the visible marks inside the main density mass of the idealized curve.
This would result in overestimation of spread for all visualization types except for boxplots, where covering the central rectangle---which only contains 50\% of the data---result in underestimation.

\end{itemize}

\begin{figure*}[htb]
    \centering
    \subfloat[Mean estimation.]{
        \includegraphics[height=6cm,page=1]{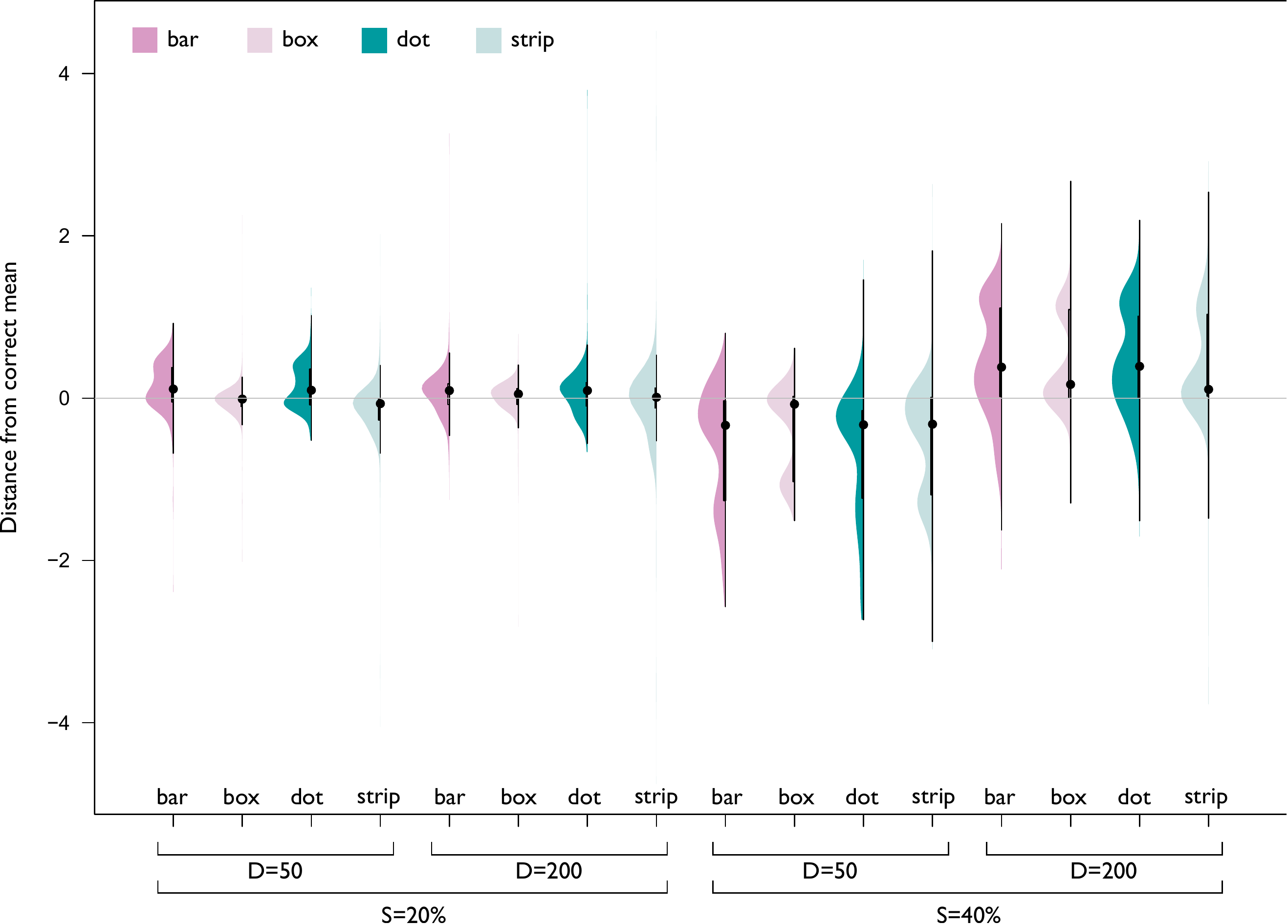}
    }
    \subfloat[Standard deviation estimation.]{
        \includegraphics[height=6cm,page=1]{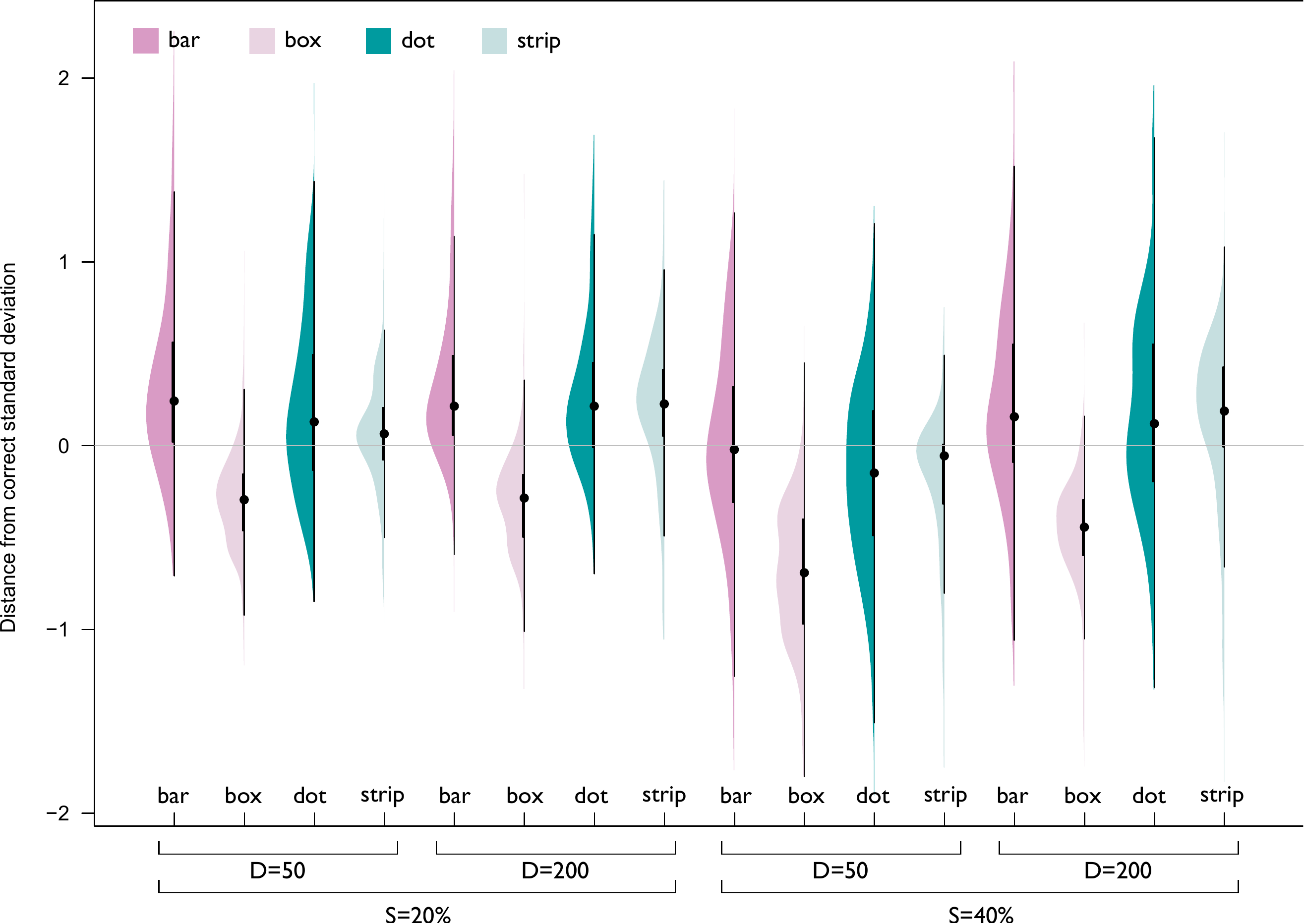}
    }
    \caption{\textbf{Detailed performance per visualization.} 
    Distribution of mean (left) and standard deviation (right) error in all trials organized by spread $S$, data size $D$, and visualization $V$.
    }
    \label{fig:perform-halfeye}
\end{figure*}

\begin{figure*}[htb]
    \centering
    \subfloat[Bar, S=20\%, D=200, r=a.]{
        \includegraphics[width=0.24\linewidth,page=5]{plots/umbrella-all}
    }
    \subfloat[Dot, S=20\%, D=200, r=a.]{
        \includegraphics[width=0.24\linewidth,page=6]{plots/umbrella-all}
    }
    \subfloat[Box, S=20\%, D=200, r=a.]{
        \includegraphics[width=0.24\linewidth,page=7]{plots/umbrella-all}
    }
    \subfloat[Strip, S=20\%, D=200, r=a.]{
        \includegraphics[width=0.24\linewidth,page=8]{plots/umbrella-all}
    }\\
    \subfloat[Bar, S=40\%, D=200, r=a.]{
        \includegraphics[width=0.24\linewidth,page=1]{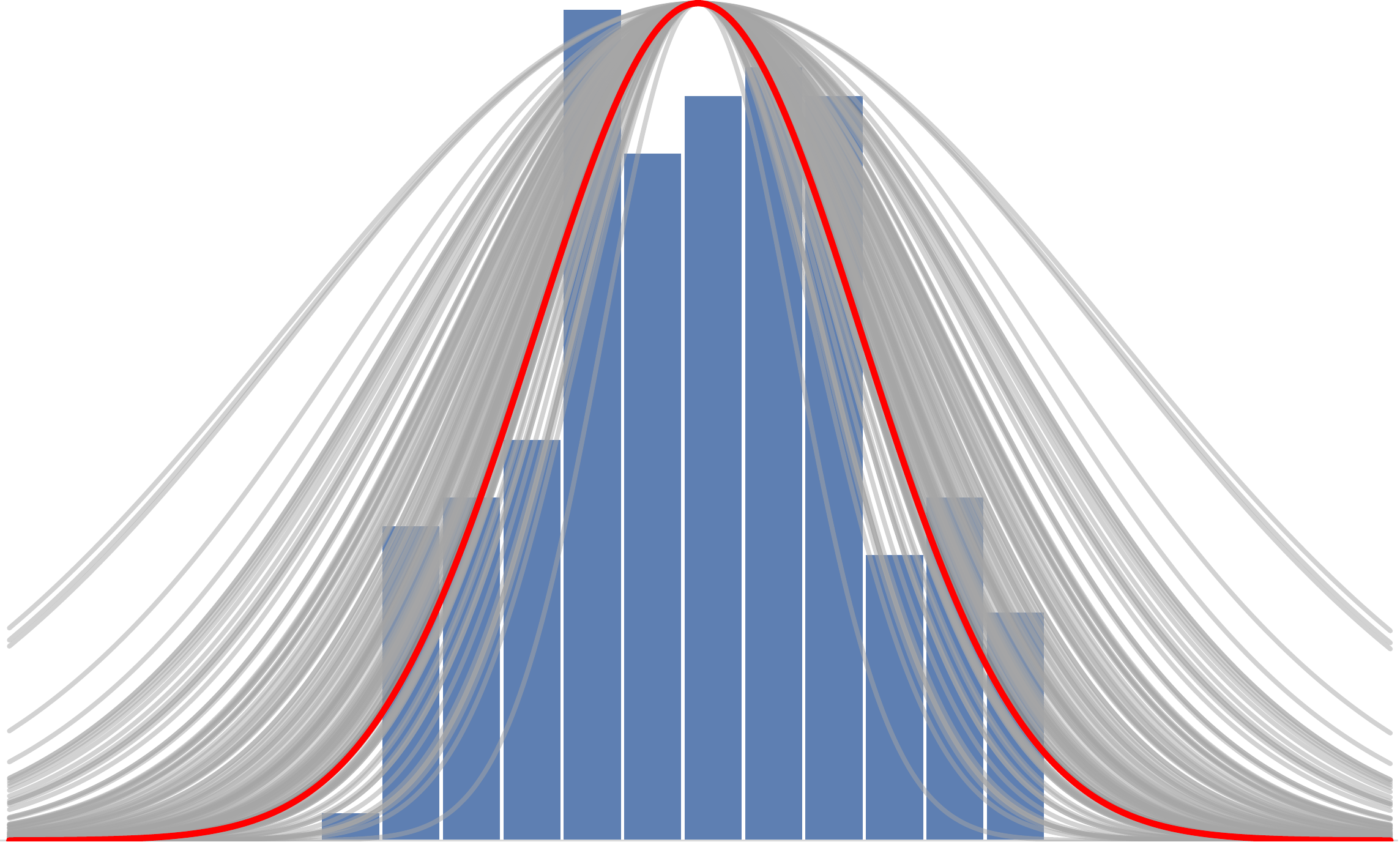}
    }
    \subfloat[Dot, S=40\%, D=200, r=a.]{
        \includegraphics[width=0.24\linewidth,page=2]{plots/umbrella-all}
    }
    \subfloat[Box, S=40\%, D=200, r=a.]{
        \includegraphics[width=0.24\linewidth,page=3]{plots/umbrella-all}
    }
    \subfloat[Strip, S=40\%, D=200, r=a.]{
        \includegraphics[width=0.24\linewidth,page=4]{plots/umbrella-all}
    }
    \caption{\textbf{Curve fitting examples.}
    All fitted curves for two specific trials: $S$=20\%, $D$=200, rep=a (top row) and $S$=40\%, $D$=200, rep=a (bottom row).
    The curves have been centered around the mean value. 
    The red curve represents the actual Gaussian of the given data sample.
    The respective visualizations have been added to the background for context.
    }
    \label{fig:umbrella}
\end{figure*}

\begin{figure*}[htb]
    \centering
    \subfloat{
        \includegraphics[width=0.48\linewidth]{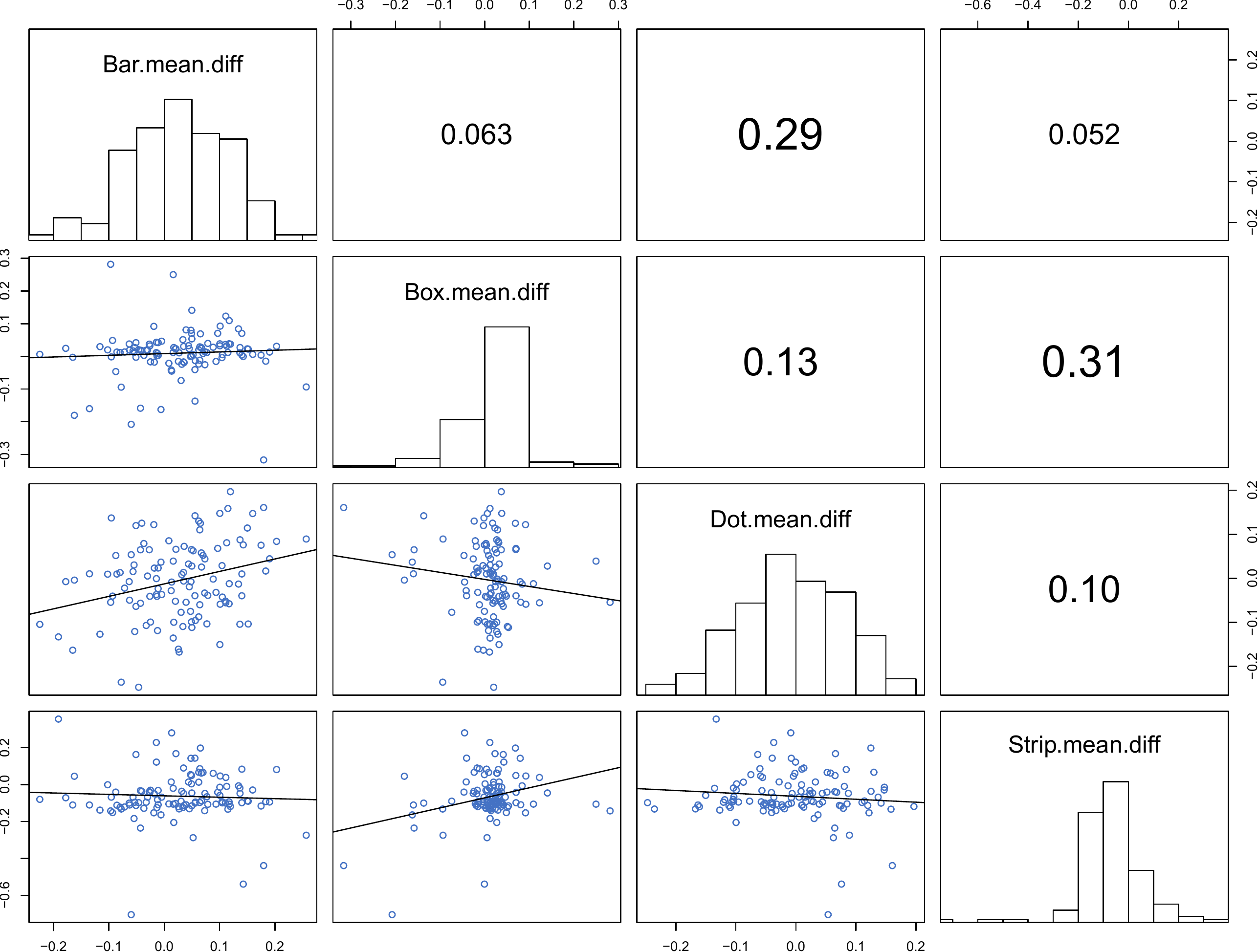}
    }
    \subfloat{
        \includegraphics[width=0.48\linewidth]{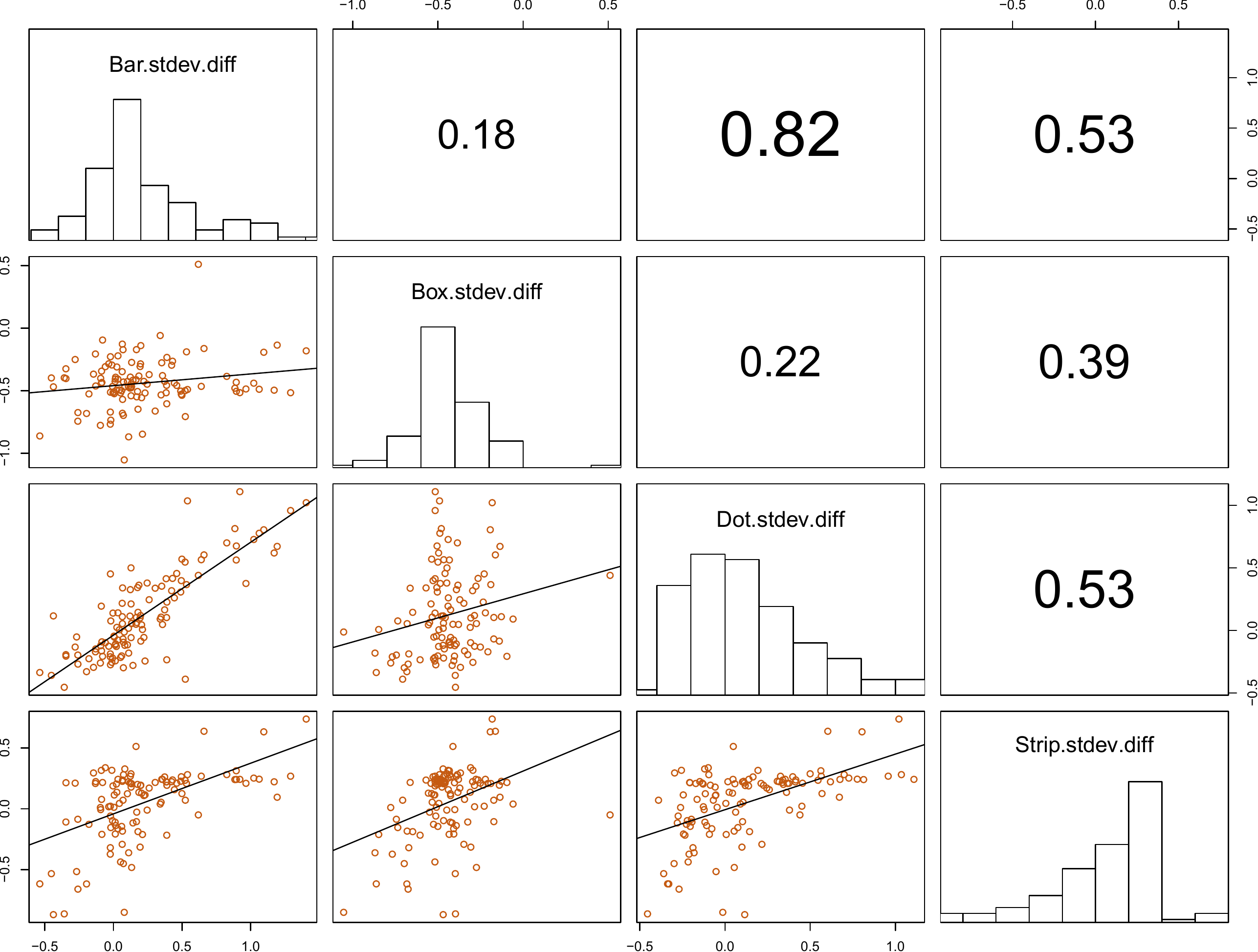}
    }
    \caption{\textbf{Performance comparison.}
    Comparing average performance across graph types of individual respondents.
    Each dot represents a respondent's average absolute error in finding the mean or standard deviation of data sets in all trials with the given graph type.
    The cells of the matrices facilitate comparisons of respondent performance on each possible graph type pairing.
    Respondent performance in finding means are in blue (left), standard deviations in range (right).
    Absolute magnitude of Pearson $r$ pairwise correlation for each pair of visualization types is given above the diagonal.
    }
    \label{fig:indperform}
\end{figure*}

\begin{figure*}[htb]
    \centering
    \colorbox{blue!30}{\parbox[t][0.25cm]{0.31\textwidth}{\centering \textsf{Mean error}}}
    \colorbox{blue!30}{\parbox[t][0.25cm]{0.31\textwidth}{\centering \textsf{Standard deviation error}}}
    \colorbox{blue!10}{\parbox[t][0.25cm]{0.31\textwidth}{\centering \textsf{Completion Time}}}
    \\
    \subfloat[]{
        \includegraphics[width=0.32\textwidth]{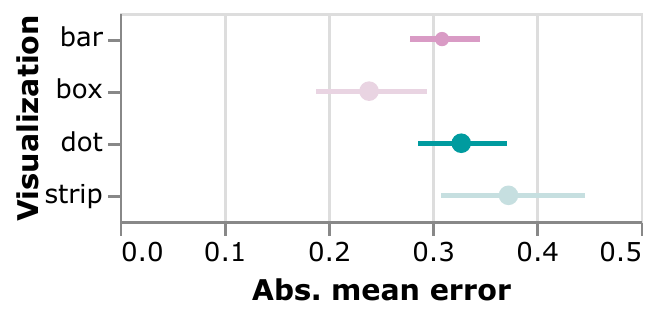}
        \label{fig:vis-meandiff}
    }
    \subfloat[]{
        \includegraphics[width=0.32\textwidth]{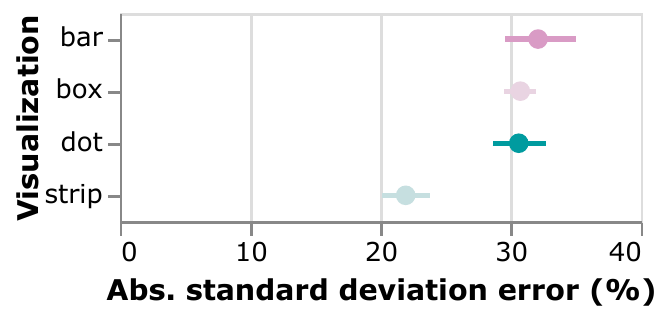}
        \label{fig:vis-sddiff}
    }
    \subfloat[]{
        \includegraphics[width=0.32\textwidth]{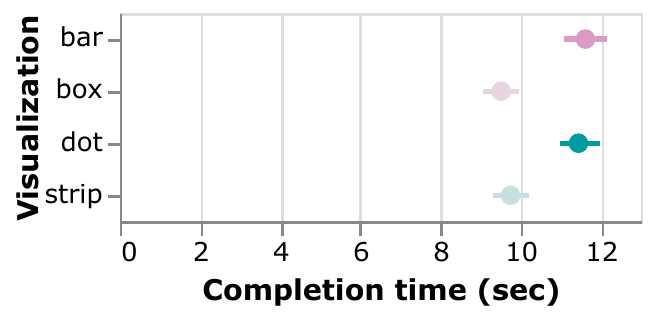}
        \label{fig:vis-time}
    }
    \\
    \subfloat[]{
        \includegraphics[width=0.32\textwidth]{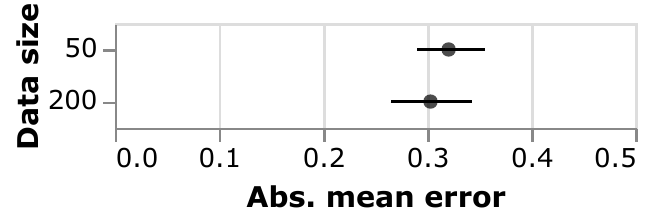}
        \label{fig:datasize-meandiff}
    }
    \subfloat[]{
        \includegraphics[width=0.32\textwidth]{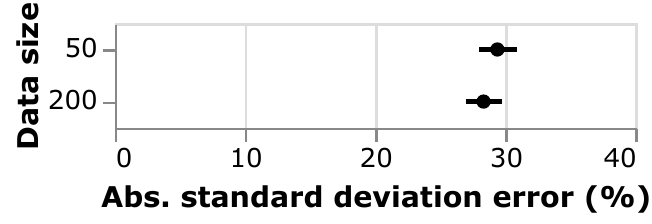}
        \label{fig:datasize-sddiff}
    }
    \subfloat[]{
        \includegraphics[width=0.32\textwidth]{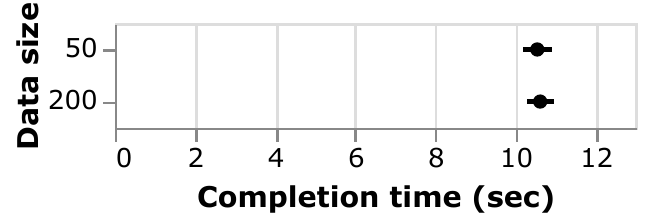}
        \label{fig:datasize-time}
    } 
    \\
    \subfloat[]{
        \includegraphics[width=0.32\textwidth]{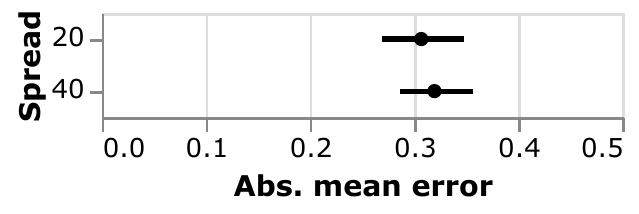}
        \label{fig:spread-meandiff}
    }
    \subfloat[]{
        \includegraphics[width=0.32\textwidth]{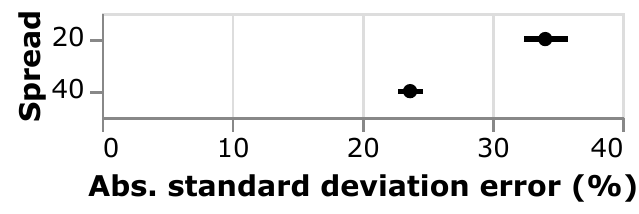}
        \label{fig:spread-sddiff}
    }
    \subfloat[]{
        \includegraphics[width=0.32\textwidth]{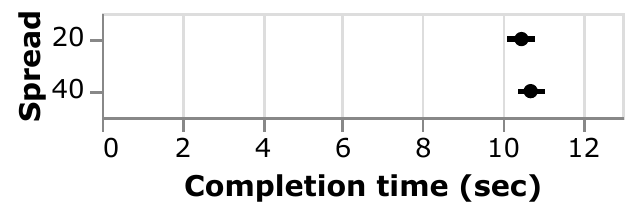}
        \label{fig:spread-time}
    } 
    \caption{\textbf{Overall performance.}
    Bootstrapped per-participant 95\% C.I.s showing the effect of Visualization $V$, Data Size $D$, and Spread $S$ on mean error, standard deviation error, and completion time.
    (Note that completion time is only included for the sake of completeness.)}
    \label{fig:all-results} 
\end{figure*}

\begin{figure}[htb]
    \centering
    \subfloat{
        \includegraphics[width=0.47\linewidth]{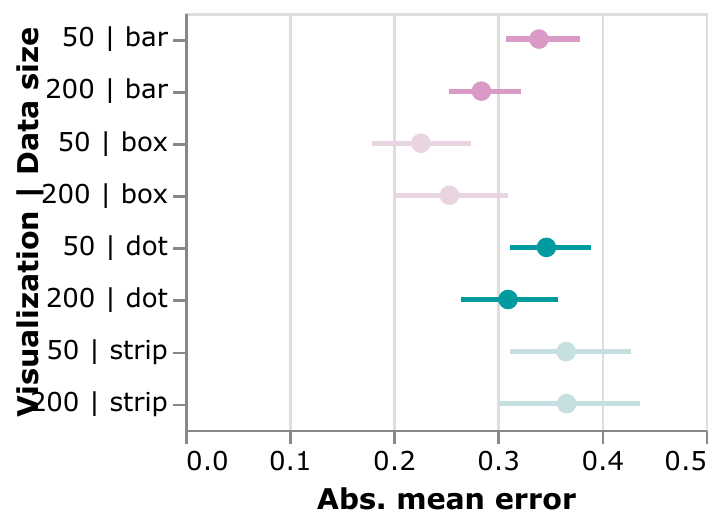}
    }
    \subfloat{
        \includegraphics[width=0.47\linewidth]{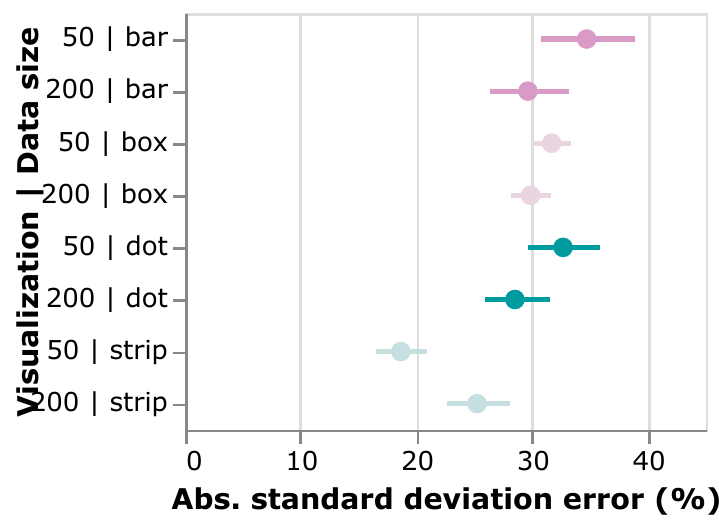}
    }\\
    \subfloat{
        \includegraphics[width=0.47\linewidth]{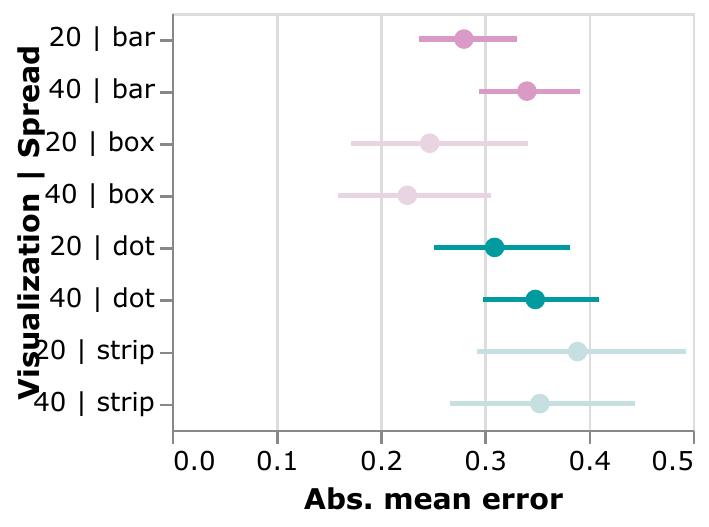}
    }
    \subfloat{
        \includegraphics[width=0.47\linewidth]{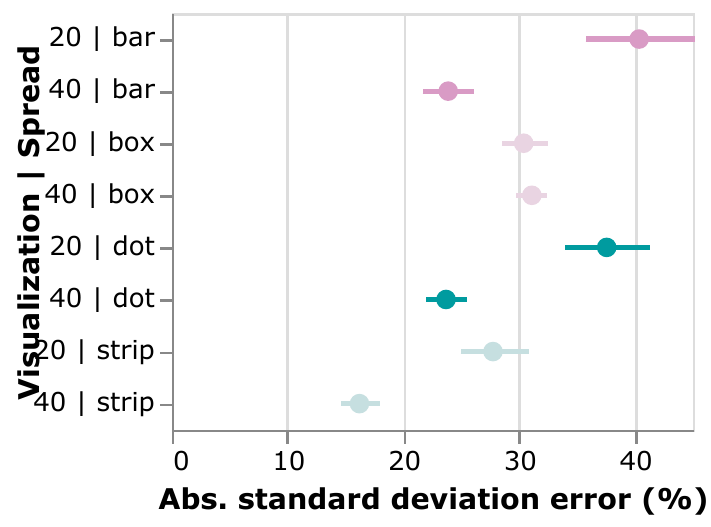}
    }
    \caption{\textbf{Interactions between factors.} Bootstrapped per-participant 95\% C.I.s showing the effect of Visualization $V$ with Data Size $D$ and Spread $S$ on absolute mean and standard deviation error.}
    \label{fig:spread-size}
\end{figure}

\section{Results}
\label{sec:results}

We analyzed the results from our study using bootstrapping~\cite{Efron1992} ($N = 1,000$ repetitions) to compute 95\% confidence intervals (CIs)~\cite{Dragicevic2016}.
\changed{Our summary calculations do not adjust for interactions between factors; see Section~\ref{sec:factor-analysis} for our analysis of interaction effects.}
We report effect sizes based on these intervals.

We collected data from 146 participants who completed 48 trials for a total of 7,008 individual trials.
After discarding the 19 participants who failed the four attention trials, we were left with 127 participants.
Upon inspecting the data, we found that an additional 10 participants appeared to have misunderstood the curve fitting task for an entire block or more of the experiment.
More specifically, these participants had moved the position of the curve to fit the mean, but had not changed the width of the curve to fit the spread.
We speculate that this problem arose because our training trials failed to require respondents to adjust the spread.
In retrospect, we should have forced participants to change the spread of a curve before they were able to submit a trial (we only forced them to interact with the slider in some way).
Since we are unable to assess the impact of this misleading training, we opted to remove the data from all of those 10 participants from our analysis.
This means we ended up with a total of 117 participants.

Based on our preregistration, we eliminated outlier trials (not participants) with a completion time higher than $3\sigma$; this removed a total of 79 trials (i.e., 1.3\% of all trials).
We assume these trials represent situations when the participant was interrupted mid-trial; most of these lasted for hundreds of seconds (the maximum was 698 seconds).
We argue that eliminating such trials based on completion time is valid both because (a) data collection using online crowdsourcing is much less controlled than in laboratory settings, thus requiring accommodations due to participant inattention, latency, and external interruptions~\cite{Heer2010}, and (b) none of our hypotheses are based on completion times.

The final dataset, after removing outliers, had 5,527 trials.
The overall average completion time was 10.5 (s.d.\ 7.99) seconds.
The overall absolute average mean error was 11.2\% (s.d.\ 25.2\%).
The overall absolute average standard deviation error was 28.9\% (s.d.\ 30.4\%).
Below we analyze participant performance and then go into detail on the characteristics of the different factors.

\subsection{Averages and Individual Analysis}

Figure~\ref{fig:perform-halfeye} shows the data distribution for all combinations of spread, data size, and visualization type for both mean estimation error and standard deviation error. 
There are a few clear trends visible.
There is a tendency for several visualizations to cause participants to overestimate the standard deviation.
The exception is boxplots, which appear to instead yield underestimation.
This overestimation is especially clear in Figure~\ref{fig:umbrella}, which shows centered fitted curves for two specific trials chosen as representative examples of ``tight'' (peaked) and ``loose'' (flatter) distributions.

\changed{Figure~\ref{fig:indperform} compares performance for individuals.
Each dot represents a respondent's average error across all trials for that graph type.
Outliers larger than $3\sigma$ have been removed from this data.
Respondent performance in estimating standard deviations shows considerable variation.
Absolute Pearson $r$ correlation values in performance between pairs of visual representations are also shown in the figure.
For absolute mean error, boxplots and strip plots appear most closely correlated ($|r| = 0.31$), closely followed by dotplots and bar histograms ($|r| = 0.29$).
On the other hand, performance on histograms does not appear to correlate well with performance on strip plots or boxplots ($|r| < 0.13$).}


For absolute standard deviation error, absolute correlations are much higher---for example, bar histograms and dotplots are highly correlated ($|r| = 0.82$); dotplots vs.\ strip plots and bar histograms vs.\ strip plots also exhibit correlation (both $|r| = 0.53$).

\begin{table}[htb]
    \centering
    \caption{\textbf{Effect sizes.}
    Absolute mean error and absolute standard deviation error (\%) for Visualization types $V$.
    \changed{The 95\% confidence intervals were calculated using bootstrapping; see Section~\ref{sec:results}.}
    }
    \label{tab:effect-sizes}
    \begin{tabular}{lrrrc}
        \toprule
        \textbf{Visualization $V$} &
        \textbf{mean} &
        \textbf{\changed{95\% CIs}} &
        \textbf{s.d.} &
        \textbf{Cohen's $d$}
        \\
        \midrule
        \textit{Absolute mean error:}\\
        -- Bar histogram & 0.312 & \changed{[0.29, 0.34]} & 0.480 & $-0.001$\\
        -- Boxplot & 0.242 & \changed{[0.20, 0.28]} & 0.713 & $-0.140$\\
        -- Dotplot & 0.328 & \changed{[0.30, 0.36]} & 0.581 & $0.031$\\
        -- Strip plot & 0.367 & \changed{[0.33, 0.41]} & 0.833 & $0.110$ \\
        \midrule
        \textit{Absolute s.d.\ error (\%):}\\
        -- Bar histogram & 32.2 & \changed{[30.2, 34.4]} & 39.3 & $0.148$\\
        -- Boxplot & 30.7 & \changed{[29.9, 31.7]} & 17.1 & $0.082$\\
        -- Dotplot & 30.6 & \changed{[28.9, 32.5]} & 34.6 & $0.078$\\
        -- Strip plot & 21.9 & \changed{[20.6, 23.1]} & 24.6 & $-0.308$\\
    \bottomrule
    \end{tabular}
\end{table}

\subsection{Analysis by Characteristics of Factors}
\label{sec:factor-analysis}

The first three rows of Figure~\ref{fig:all-results} summarizes performance for all three measures based on Visualization $V$, Data Size $D$, and Spread $S$ using 95\% confidence intervals (calculated using bootstrapping as discussed above).
Furthermore, Table~\ref{tab:effect-sizes} summarizes the effect sizes for absolute mean error and standard deviation error (\%).
\changed{We used the classic Cohen's $d$ formulation, i.e., one independent of the experimental design and thus amenable to comparison with other experiments.}
For Visualization (the first row), error in estimating the mean was lower for boxplots (absolute mean error=0.24, Cohen's $d=-0.14$) compared to the other chart types.
Strip plots had slightly worse performance (absolute mean error=0.37, $d=-0.11$), but were comparable in performance to bar and dotplots.
Participants were considerably more accurate in estimating standard deviation with strip plots compared to others (absolute percentage s.d.\ error=0.21, Cohen's $d=-0.31$).
Bar charts were the least accurate for estimating standard deviation (absolute percentage s.d.\ error=0.32, Cohen's $d=-0.15$).

The second row in Fig.~\ref{fig:all-results} summarizes measures for the data size.
Although larger data size was associated with better performance, these effects were small both for mean error (0.32 for $D = 50$, 0.30 for $D = 200$, Cohen's $d = 0.02$) and standard deviation error (29.4\% for $D = 50$, 28.4\% for $D = 200$, Cohen's $d = 0.03$).

On the final row of Figure~\ref{fig:all-results}, we see the same data for spread.
Here, while larger spread yields minimally higher mean error (0.307 for 20\%, 0.317 for 40\%, Cohen's $d = 0.0142$), it was associated with a larger relative performance gain for standard deviation (34.1\% for 20\%, 23.7\% for 40\%, Cohen's $d = 0.341$).

Finally, we present the interactions between spread and data size with visualization in Figure~\ref{fig:spread-size}.
Some interesting observations:

\begin{itemize}
\item \textit{For interactions between data size and visualization}, absolute mean error appears to decrease with larger data size. 
The exception is boxplots, which have consistently low absolute mean error for both 50 and 200 items.
As for absolute standard deviation error, the same trend recurs---the error appears to decrease with higher data sizes.
Strip plots exhibit consistently lower error than the other techniques for both data sizes.

\item \textit{For interactions between spread and visualization}, the absolute mean error appears to increase with higher spread; however, this effect does not persist for boxplots, which exhibit similar absolute mean error for both levels of spread.
Furthermore, for this same interaction, all visualizations yield lower standard deviation error for higher (40\%) compared to lower (20\%) spread, again except for boxplots; boxplots have largely unchanged standard deviation error for both conditions.


\end{itemize}

\begin{figure}[htb]
    \centering
    \includegraphics[width=\linewidth]{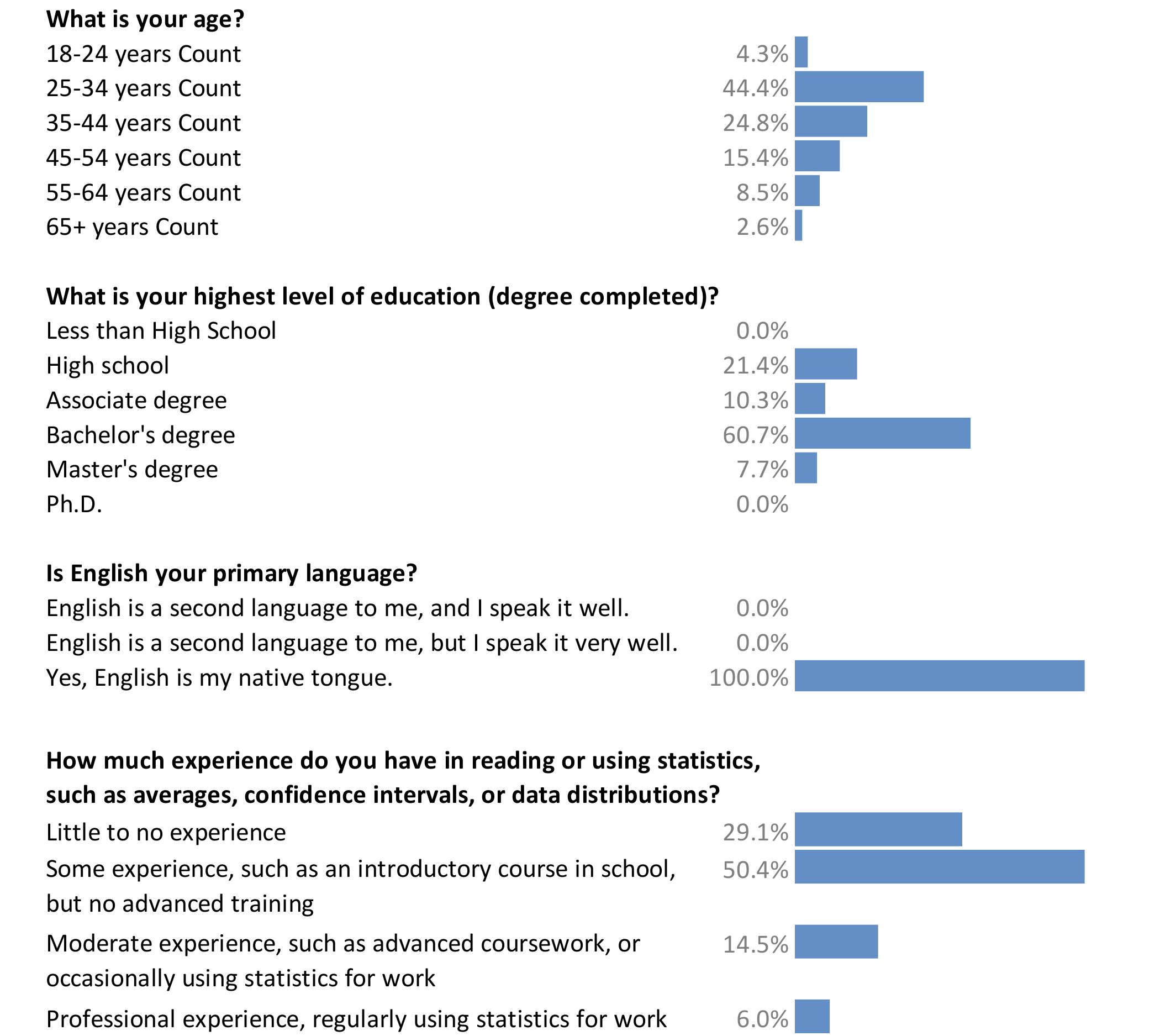}
    \caption{\textbf{Demographics.} Summary of participant demographics.}
    \label{fig:demo}
\end{figure}

\subsection{Demographics and Participant Feedback}

Figure~\ref{fig:demo} summarizes results from our demographics survey.
These are more or less consistent with prior attempts at understanding the Amazon Mechanical Turk population~\cite{Ross2010}. 
Importantly, very few of our respondents (8) had professional statistics experience, and only about 1 in 7 had moderate statistics experience.

Free-form feedback was provided by 37 out of the 117 participants.
The feedback was generally positive, with several participants noting that the task was fun and engaging.
Wrote participant P21812, \textit{``It was almost gamified, like something you'd find on a tablet or phone.''}
Similarly, \textit{``This was fun! I kept wanting to fit everything in under the bell curve, though the instructions didn't state that.''} (P51429)
Many other participants also noted that fitting curves was challenging:
\textit{``Much of the data was not in a normal distribution!''} (P15279),
\textit{``Some of those were tough, just trying to eyeball them.''} (P28558), and
\textit{``Not sure exactly how these were supposed to fit, some were too weird.''} (P49775)

Finally, participant P21883 had a more general comment:
\textit{``I took high school statistics but I don't remember estimating the curves this way.  I just remember doing tons of calculations.''}
P43443 went even deeper: \textit{``I think I would have had better luck at proving string theory than doing this exercise with any degree of accuracy.''}
Fortunately, our results disagree.

\subsection{Deviations from the Preregistration}
\label{sec:deviations}

In our preregistration, we stated that we would collect data from 100 participants.
For the actual study, we recruited a total of 150 participants (and ended up with 146 participants, and eventually 117 after filtering), which was an unintentional deviation from the preregistration. 
The cause for this deviation was miscommunication within the research team.
Though not presented here, we did a separate analysis on the first 100 respondents which indicated the same effects as the full results we have reported. 

Because of the misleading training, where participants were not required to change the spread to fit the blue curve, we also removed data for 10 participants that we deemed to have misunderstood the curve fitting task.
We classified such misunderstanding when at least one full block of 48 trials for a participant had a spread of 1.0 (the starting spread).
We provide our unfiltered data on OSF.

Additional deviations include the following:
\begin{itemize}
    \item Our spread values were said to be ``determined by pilot testing.''
    This is incorrect---we had determined the values based on experiences from the prior version of this experiment.
    
    \item Our preregistration included a fifth hypothesis: ``There will be non-uniformity in performance across individual participants.''
    This hypothesis is underspecified and ambiguous, and we opted to discard it in our analysis.
    
    \item Several of our plots and analyses---including our analysis of Pearson correlations and Cohen's effect sizes as well as the scatterplot matrix in Figure~\ref{fig:indperform}---were not included in our preregistration.
    We include them here in the interest of providing a richer analysis and reporting of our results.
\end{itemize}

\section{Discussion}

Our study gave rise to several interesting findings while confirming our basic intuitions about this task.
First we review our hypotheses in light of these results.
Then we explain our results in detail and discuss how they generalize. 
We close by discussing the overarching research vision motivating this work.

\subsection{Reviewing Hypotheses}

We find strong evidence in our results \textbf{that the estimation of means is more accurate than estimation of spread}.
Overall, the left and center columns of plots in Figure~\ref{fig:all-results} indicate that people have a much easier time estimating the average values in a dataset rather than its spread, regardless of visualization technique, data size, or spread.
This supports our first hypothesis.

\changed{We observed that errors in estimating both mean and standard deviation decrease as data size increases for both bar and dotplot histograms.
However, for other visualizations, there appears to be no such effect, and for strip plots, the error actually increases as data size increases.
While this could be said to partially support our second hypothesis, we think it rather suggests that this hypothesis---more data equals better estimation---is overly simplistic.
For example, boxplots do not even encode data size, and strip plots seem to not scale well as the number of strips increases.}

In keeping with our third hypothesis, the top row of Figure~\ref{fig:all-results} illustrates some non-uniformity in performance across visualization types.
As per our expectations, \textbf{boxplots yielded the lowest error and strip plots the highest for estimating the mean.}

Our fourth hypothesis dealt with the assumption that participants would seek to fit as many of the visible data points in the sample as possible under the curve.
We call this an ``umbrella effect,'' as if people are protecting the data from rain.
If so, this would result in systematic overestimation of standard deviation for visualizations other than boxplots.
Figures~\ref{fig:perform-halfeye}, \ref{fig:umbrella}, and \ref{fig:indperform} show that people routinely overestimated the spread for bar and dot histogram plots, but this phenomenon does not appear to extend to boxplots and strip plots.
This suggests the presence of the umbrella effect for these plots, supporting the fourth hypothesis.
It may also apply for the boxplot, as the underestimation may indicate participants are fitting the whiskers, which would exclude some data, or even boxplot rectangle, which only contains 50\% of the data.

\subsection{Explaining the Results}

We note that the error in estimating means does not increase with higher spread $S$; basically, higher spread derived from more noise in the sample should make it harder to determine an accurate mean for the histogram.
Similarly, we our results indicate the error is unchanged as the data size $D$ increases. 
We would have expected that an increasing number of data points would build a fuller picture of a distribution, easing the perceptual task of finding the midpoint.
We see some evidence of this, as discussed at the end of Section~\ref{sec:factor-analysis}, but it is not nearly as strong as expected.

Our results on standard deviation error are also noteworthy.
First of all, the impact of data size $D$ on this error is small.
This would usually be interesting as one would guess that increasing the number of items would yield better accuracy because the sample becomes more regular.
However, we controlled the relative noise of the samples when we selected CV levels.
On the other hand, we do see that standard deviation error decreases as the spread $S$ increases.
This is counterintuitive, but may possibly be explained by the aforementioned ``umbrella effect.''
Figure~\ref{fig:umbrella} tells this story.
This is further supported by our results that indicate that most trials were overestimates, i.e., with standard deviations larger than necessary.
Results for boxplots were the reverse, showing a strong negative bias in errors of standard deviation estimation, a reverse umbrella.
We speculate that this may reflect participants attempting to fit the curve directly onto the figure.  
The excellent results for estimating means with boxplots may simply reflect how on an individual basis we cannot compete with the precision offered by automated calculations, despite our innate ability to identify the mean of a distribution.
In samples from normal distributions, the precalculated medians denoted by the center lines of boxplots are very close to the distribution mean.
The other points making up a boxplot represent similar precomputed values.
Boxplots thus present distributions with an appearance of having very little random noise.

A normal curve goes even further, being a completely noiseless and idealized representation of a data sample.
Upon reflection, it seems only reasonable that participants would have greater ease applying a normal curve to the most noise-free visualization among the four we tested.
As data visualization researchers, we tend to assume there is useful information in deviations of data from some idealized model; we look for meaning in the details.
Yet that same information---outliers, unexpected correlations, gaps in the data---can apparently act as a distraction from ''eyeballing'' traditional summary measures.   
It may be that the idealized forms traditional statistics focus upon sometimes make a poor match with our intuitions for messier, real world data.
A normal curve is a structure we impose on smaller samples whenever we calculate a mean and standard deviation, rather than an obvious fit.
This may highlight the importance of getting a sense of the data by looking at it before it is abstracted or summarized.
Yet it also argues for the importance of summary calculations that can precisely identify (often critical) measures of central tendency, since noise in the data may distract our visual capacity.

\subsection{Generalizing the Results}

Our participants included a high concentration of younger adults, with a higher ratio of university degrees than in the U.S.\ general population. 
However, the level of statistics knowledge participants report was relatively basic.
This low level of statistics training gives us confidence that the findings may generalize more broadly.

We believe that the sample sizes of our trial datasets (50 and 200 data items) represent typical sample sizes for people in many fields of study.
The smaller size does raise an issue for the applicability of these results to future research into visual inference, since below 100 cases, the Student's t-distribution is not totally equivalent to the normal distribution~\cite{Student1908}.
However, at $n = 50$, the two curves are similar, and we expect that the challenge-to-fit presented by the noisiness of the small sample would be the more important effect.   

The visualizations we tested varied in their degree of aggregation, spanning the full range from zero aggregation (strip plots), to partial spatial aggregation (dot histograms), moderate aggregation (bar histograms), and finally a high degree with boxplots.
We believe that other techniques for visualizing distributions, for example, violin plots or density plots, might be similarly scored along this dimension. By so doing, future researchers might use the results of this study to inform performance predictions for these or other graphics.
However, other techniques may exhibit entirely different behavior.
For instance, hypothetical outcome plots (HOPs)~\cite{Hullman2015}, for example, use animation to convey uncertainty, and so, while they are not aggregated, rely on spatial memory and other visual mechanisms that we did not investigate in this study.

\changed{As shown in our study of interactions (Section~\ref{sec:factor-analysis}), aggregation is a double-edged sword.
While other visualization types showed variation in performance in estimating standard deviation as the data size and spread fluctuated, boxplots had consistent performance across these properties.
Yet, this \textit{consistent} error was relatively \textit{high}: boxplots were relatively robust in appearance regardless of the distribution, but this simplicity could hide or mislead viewers about more fine-grained properties of the distribution.
}

\subsection{Limitations}

Even though the results presented here represents the second incarnation of our study, our experiment has some weaknesses.
For one thing, we made several deviations from our preregistration (Section~\ref{sec:deviations}), and a few of our analyses were also not explicitly detailed in the preregistration.
In out future work, we will endeavor to be more precise and prescriptive even in its planning stages.

Our training trials primed participants to answer our questions by showing a prefitted idealized curve on top of a data distribution for each visualization type.
However, it is conceivable that such training teaches people less about fitting curves to data than fitting curves to a specific visualization.
This is an entirely fair point, and is essentially also the purpose of our experiment: to understand how well different visualizations convey the mean and standard deviation of a normal data sample.
Still, we tried to avoid training people to mechanically fit curves using a prescribed pattern by only giving participants a single training trial per visualization type.

Furthermore, our study only involved normally distributed data.
This is a clear limitation to our experiment, and more work is needed to chart these waters in the future.
In addition, we opted not to include the number and configuration of bin sizes, which have been shown to be important factors in histogram design~\cite{Correll2019},
in our experiment to keep the size of the experimental design manageable, and instead held the number of bins (50) and their size ($20.0/50 = 0.40$ per bin) constant.
This is another limitation of our work, as two of our techniques---bar and dot histograms---clearly are affected by this choice, whereas the other two---strip plots and boxplots---are not.
While we think that the bin sizes were more or less appropriate given the dataset properties, this is nevertheless an important factor that we hope will be studied in the future.

Finally, our work here is focused on fitting curves for use in classic parametric statistical tests.
You might argue that basing this work on such classic tests is counterproductive in light of more modern tests potentially better suited to graphical inference.
While we do not necessarily disagree with this sentiment, we note that (1) fitting normal distributions on discrete samples is a common task for many basic statistical operations---not the least the ability to simply characterize a distribution---and that (2) t-tests and other parametric statistical tests are still fundamental statistical tools.

\subsection{Future Work}

The overarching vision motivating our work in this paper is to lower the threshold of using real statistical methods so that anyone with little mathematical or statistical background can use them.
This vision is inspired by findings from perception and visualization (e.g.,~\cite{Albers2014, Correll2014, Gleicher2013}) that shows how  appropriate visual representations can enable even sophisticated statistics.

Our work in this paper represents one step towards such a vision.
Our findings suggest that people with minimal training can gain statistically useful information from visual displays.
More importantly, our findings show that people are \textit{consistent}; they tend to overestimate or underestimate the standard deviation depending upon the visualization used.
This gives us hope that we can design visual interventions to overcome these biases as the following step, and suggest fruitful avenues of future empirical work such as exploring more complex scenarios like comparisons of multiple distributions or constructing intervals.

Looking to the future, we view the simple estimates of statistical moments in single distributions as an important and necessary stepping stone for the project of improving ``visual statistics''~\cite{Correll2015} as well as charting the limits and potential pitfalls in ``graphical inferences''~\cite{Buja2009}.
For example, the heterogeneity of performance across visualization types in our experiment suggests that a space for potential interventions to improve an audience's grasp of the properties of a distribution, such as ``hybrid'' visualizations that combine multiple univariate visualizations~\cite{Correll2019}.

\section{Conclusion}

We have presented results from a preregistered crowdsourced user study on fitting normal curves onto more or less noisy data samples visualized using different representations. 
We find that our participants are good (approximately 12\%-13\% error) at accurately determining the mean of a data sample and that they can determine the standard deviation of the sample with 28\%-29\% error.
We think that these results are encouraging for visualization designers that rely on their audiences having a good grasp of mean and spread---especially for cases where these designers rely on implicit estimates or satisficing strategies~\cite{Kale2021}.



\bibliographystyle{IEEEtran}
\bibliography{fitting-curves}

\begin{IEEEbiography}[{\includegraphics[width=1in,height=1.25in,clip,keepaspectratio]{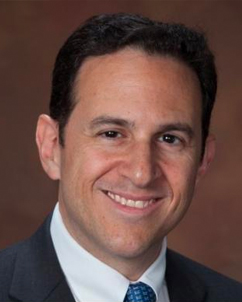}}]{Eric Newburger}
received the masters degree in Applied Economics in 1995 from the University of Wisconsin -- Madison, in Madison, WI, USA.
He is a Ph.D.\ candidate and adjunct lecturer in the College of Information Studies, University of Maryland, College Park in College Park, Maryland, USA. 
He is also a student member of the Human-Computer Interaction Laboratory (HCIL) at UMD.
\end{IEEEbiography}

\begin{IEEEbiography}[{\includegraphics[width=1in,height=1.25in,clip,keepaspectratio]{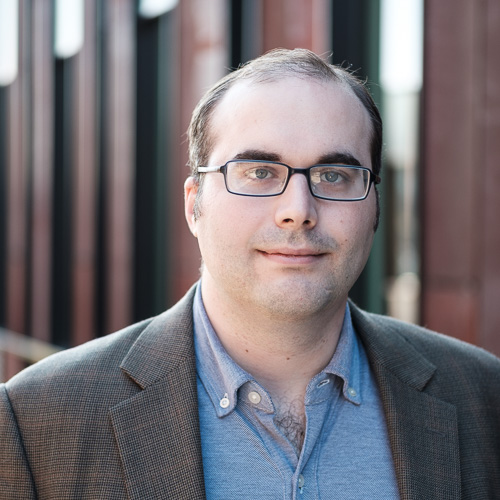}}]{Michael Correll}
received the Ph.D.\ degree in 2015 from the University of Wisconsin -- Madison in Madison, WI, USA.
He is a Senior Research Scientist at Tableau Research as part of Tableau Software in Seattle, WA, USA.
His research interests include data ethics, communicating statistics to mass audiences, and investigating biased or misleading data visualizations.
\end{IEEEbiography}

\begin{IEEEbiography}[{\includegraphics[width=1in,height=1.25in,clip,keepaspectratio]{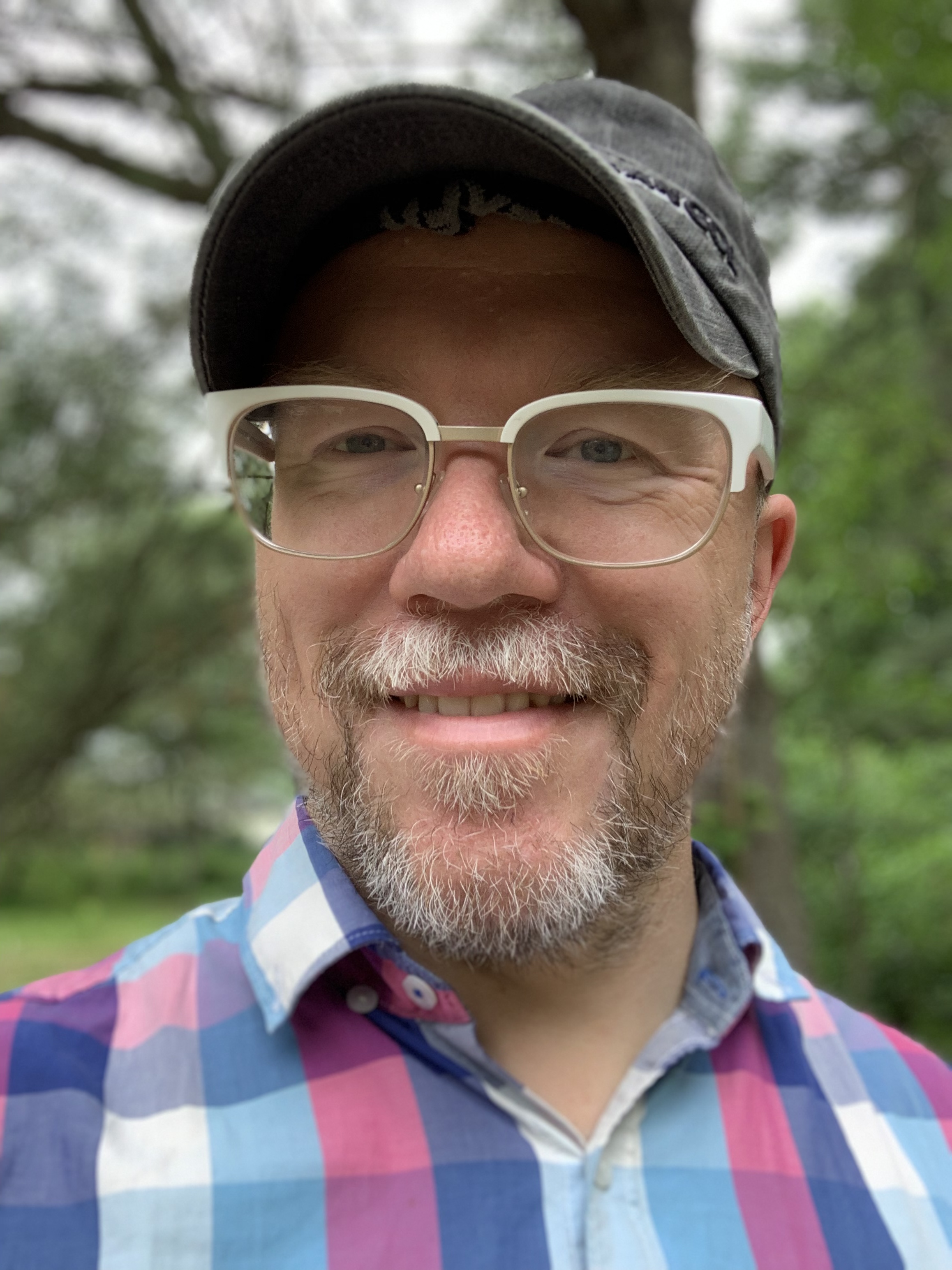}}]{Niklas Elmqvist}
received the Ph.D.\ degree in 2006 from Chalmers University of Technology in G\"{o}teborg, Sweden.
He is a professor in the College of Information Studies, University of Maryland, College Park in College Park, MD, USA. 
He is also a member of the Institute for Advanced Computer Studies (UMIACS) and formerly the director of the Human-Computer Interaction Laboratory (HCIL) at UMD.
He is a senior member of the IEEE and the IEEE Computer Society.
\end{IEEEbiography}

\vfill

\appendices
\onecolumn
\pagenumbering{arabic}
\setcounter{page}{1}
\setcounter{section}{0}
\setcounter{figure}{0}
\clearpage
\newpage

\begin{center}
  \huge{\textsf{Fitting Bell Curves to Data Distributions using Visualization}}
  
  \Large\textsf{Supplementary Materials}
\end{center}

\section{Additional Material}

All of the user study materials have been made available on our OSF website: \url{https://osf.io/r3jpn/}

\section{Study Instructions}

We are asking you to fit a smooth \textit{bell curve} to a noisy data \textit{sample}.
The sample has been randomly drawn from a source that is itself in the shape of a bell curve; imagine that it represents the heights of a large group of people, where some individuals are tall or short, but most are in the middle.
However, since the sample is small and randomly drawn, its shape can be a little rough.
Your task is to fit the bell curve on top of the data sample by adjusting its center and degree of spread, as a best guess of the approximate shape of the source population.

\begin{center}
    \includegraphics[width=0.7\linewidth]{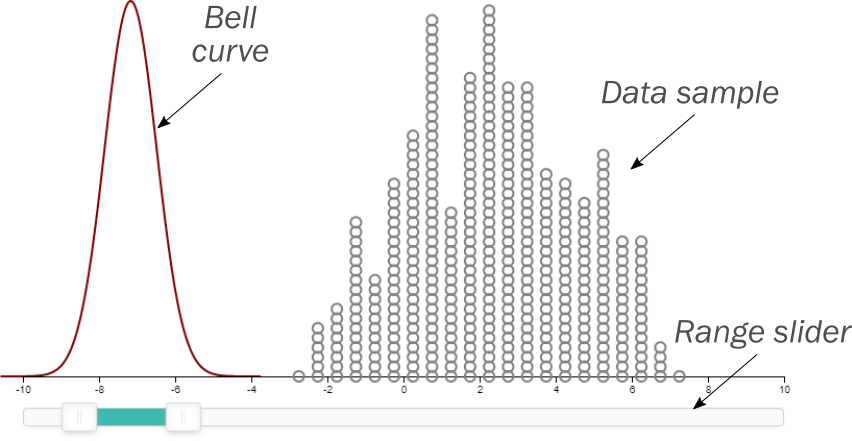}\\
    \includegraphics[width=0.7\linewidth]{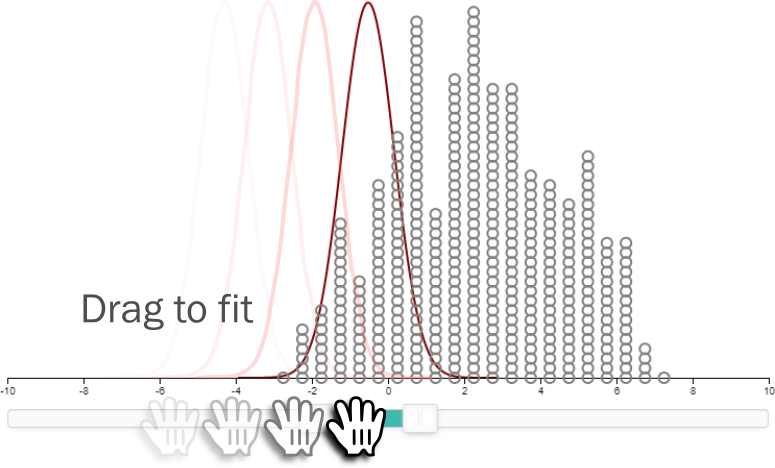}\\
    \includegraphics[width=0.7\linewidth]{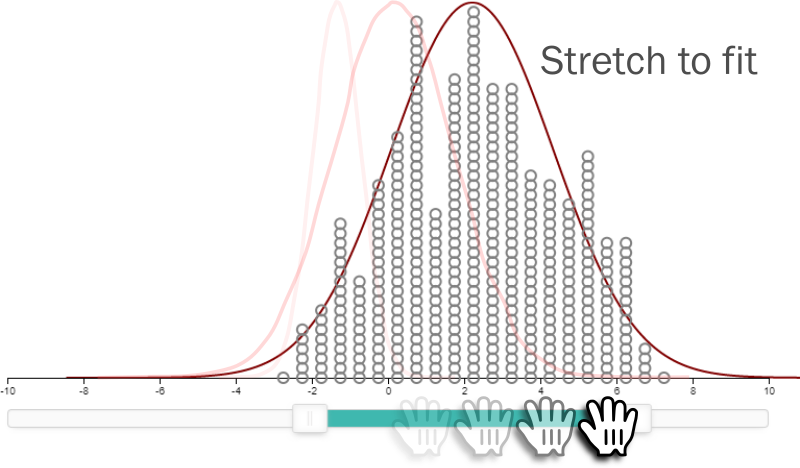}
\end{center}

The gray dots here represent each sample point.
As you can see from the image, the goal is to ``fit'' the red curve on top of the gray dots.
This means two things: (1) The curve should be \textbf{centered} on top of the gray dots.
The curve should be \textbf{stretched} so that it approximates the shape of the dots.

When fitting, the goal is \textbf{not} that all dots are inside the curve.
Instead, you want to find the best approximate fit between the dots and the curve.
Once you are satisfied, click ``\textbf{submit}.''




\section{Additional Visualizations}

\begin{figure*}[htb]
    \centering
    \includegraphics[width=0.75\linewidth]{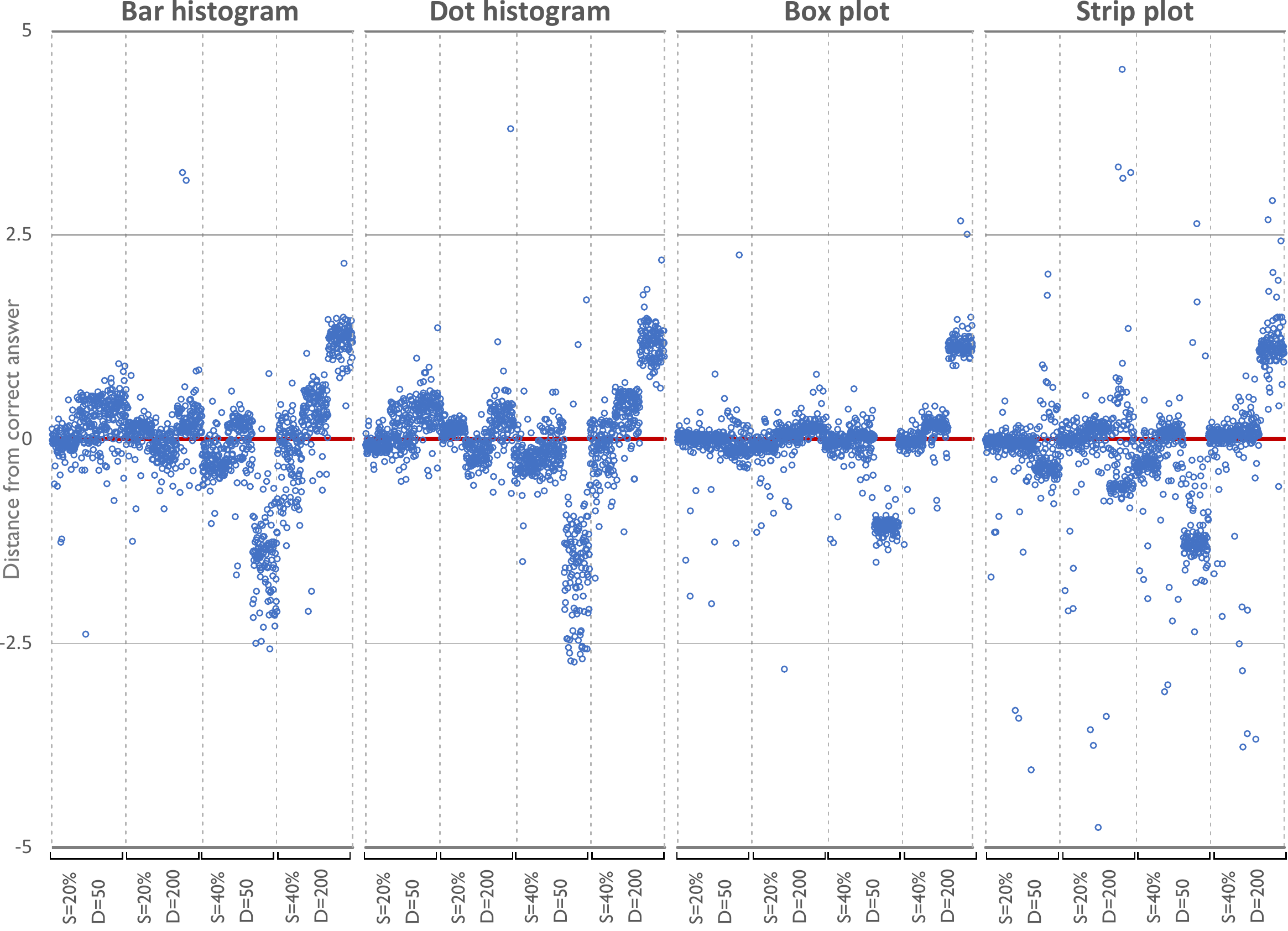}
    \caption{\textbf{Mean error performance.}
    Distance between true values and individual participant estimates of dataset means.
    The red line denotes zero distance (correct answer).
    Each region for a specific spread $S$ and dataset size $D$ represent sets of trials, with the same datasets used for each set).
    }
    \label{fig:mean-perf}
\end{figure*}

\begin{figure*}[htb]
    \centering
    \includegraphics[width=0.75\linewidth]{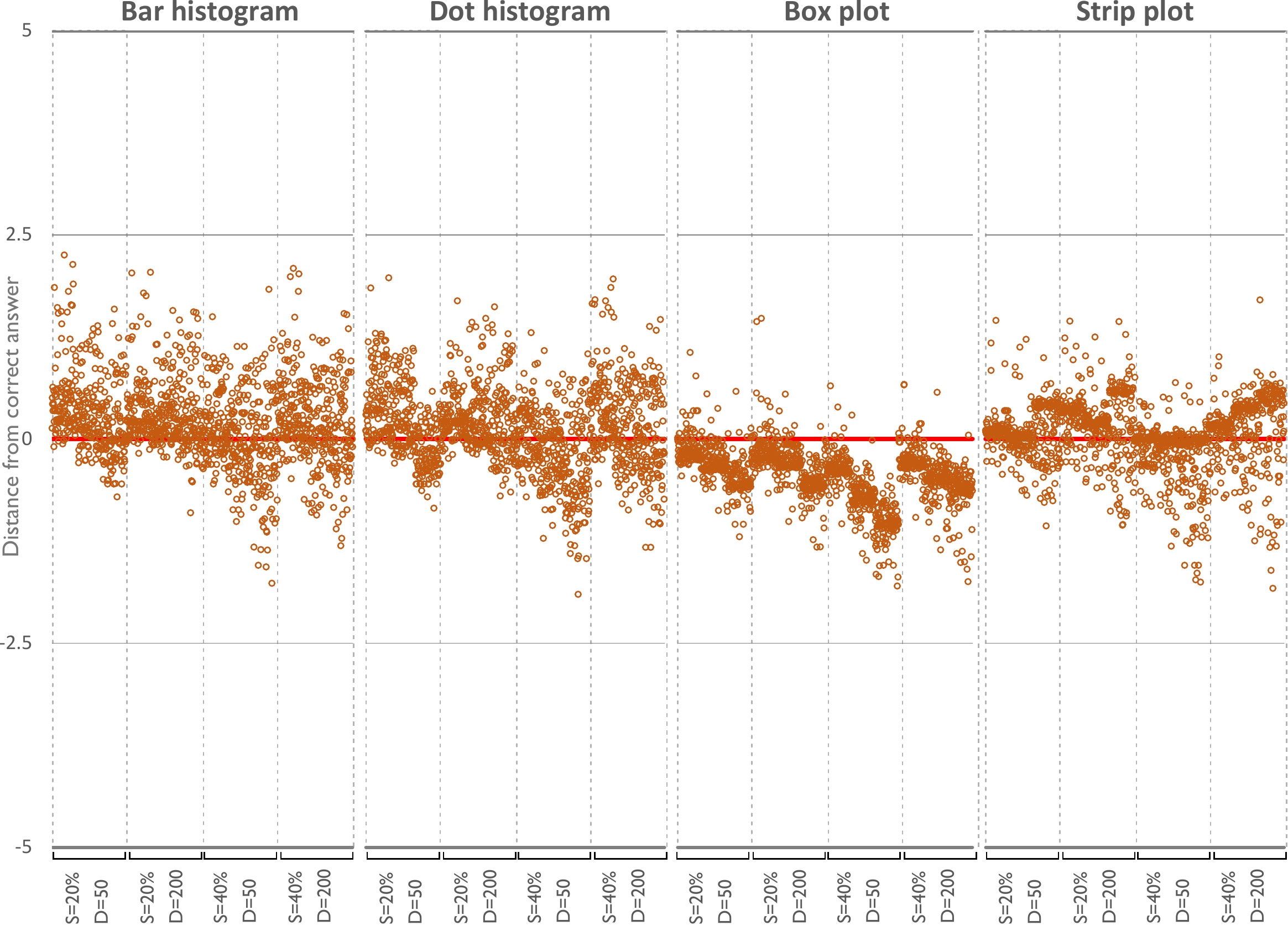}
    \caption{\textbf{Standard deviation error performance.}
    Distance between true values and individual participant estimates of dataset standard deviations.
    The red line denotes zero distance (correct answer).
    Each region for a specific spread $S$ and dataset size $D$ represent sets of trials, with the same datasets used for each set).
    }
    \label{fig:sd-perf}
\end{figure*}

Figures~\ref{fig:mean-perf} and \ref{fig:sd-perf} show individual participant performance for mean error and standard deviation error, respectively.

\end{document}